\documentclass[a4paper,fleqn]{cas-sc}
\usepackage{float}
 \usepackage[numbers]{natbib}
\usepackage{amsmath}
\DeclareMathOperator*{\argmax}{arg\,max}

\usepackage{array}
\usepackage{graphicx,epstopdf}
\usepackage{tabularx}

\usepackage{booktabs}
\def\tsc#1{\csdef{#1}{\textsc{\lowercase{#1}}\xspace}}
\tsc{WGM}
\tsc{QE}
\tsc{EP}
\tsc{PMS}
\tsc{BEC}
\tsc{DE}
\newtheorem{theorem}{Theorem}

\newtheorem{cor}{Corollary}
\newdefinition{rmk}{Remark}
\newproof{pf}{Proof}
\newproof{pot}{Proof of Theorem \ref{thm}}

\begin{document}
\let\WriteBookmarks\relax
\def\floatpagepagefraction{1}
\def\textpagefraction{.001}

\shorttitle{Novel Discrete Composite Distributions}

\shortauthors{Bowen Liu et~al.}

\title [mode = title]{Novel Discrete Composite Distributions with
Applications to Infectious Disease Data}                      
%\tnotemark[1,2]


\author[1]{Bowen Liu}[orcid=0000-0002-6918-9871]

\cormark[1]


\ead{bowen.liu@umkc.edu}

\credit{Conceptualization of this study, Methodology, Software}

\affiliation[1]{organization={University of Missouri-Kansas City},
    addressline={5000 Holmes St}, 
    city={Kansas City},
    % citysep={}, % Uncomment if no comma needed between city and postcode
    postcode={64110}, 
    state={MO},
    country={USA}}

\author[2]{Malwane M.A. Ananda}

\credit{Conceptualization of this study, Methodology}

\affiliation[2]{organization={University of Nevada, Las Vegas},
    addressline={4505 S Maryland Pkwy}, 
    city={Las Vegas},
    % citysep={}, % Uncomment if no comma needed between city and postcode
    postcode={89154}, 
    state={NV},
    country={USA}}


\cortext[cor1]{Corresponding author}


\begin{abstract}
It was observed that the number of cases and deaths for infectious diseases were associated with heavy-tailed power law distributions such as the Pareto distribution. While Pareto distribution was widely used to model the cases and deaths of infectious diseases, a major limitation of Pareto distribution is that it can only fit a given data set beyond a certain threshold. Thus, it can only model part of the data set. Thus, we proposed some novel discrete composite distributions with Pareto tails to fit the real infectious disease data. To provide necessary statistical inference for the tail behavior of the data, we developed a hypothesis testing procedure to test the tail index parameter. COVID-19 reported cases in Singapore and monkeypox reported cases in France were analyzed to evaluate the performance of the newly created distributions. The results from the analysis suggested that the discrete composite distributions could demonstrate competitive performance compared to the commonly used discrete distributions. Furthermore, the analysis of the tail index parameter can provide great insights into preventing and controlling infectious diseases. 

\end{abstract}

% \includegraphics{figs/grabs.pdf}

\begin{highlights}
\item A new type of discrete distributions is created to fit the count data with heavy tails. 

\item Two special members of this family demonstrated great performance in fitting two different infectious disease data sets. 

\item Hypothesis testing procedures are developed to provide statistical inference on tail behaviors in the context of infectious diseases case counts. 

\item When fitting to the data set with excess zeros, the family can provide performance comparable to that of the well-known zero-inflated models.  
\end{highlights}

% Keywords

\maketitle

\section{Introduction}

\label{intro}
\par 
The statistical modeling of cases or deaths of infectious diseases with parametric distributions has been an important research topic since the last century. While both continuous and discrete distributions were used to model the cases or deaths of infectious diseases, the discrete distributions were considered to be more proper since the number of cases or deaths are always reported as non-negative integers.  Thus, fitting the number of cases or deaths to the distributions such as Poisson distribution or negative binomial distribution is a common practice in infectious disease modeling. In the past, these commonly used distributions provided good performance when fitting to the infectious disease data, and they were usually well-accepted as the common distributions to fit the infectious diseases cases or deaths. 
\par  However, when the distributions of the cases or deaths are heavily right skewed with large variations, models such as Poisson or negative binomial might fail to demonstrate good performances. For example, it was observed that the extreme upper tail of the COVID-19 cases at different counties in US, followed heavy-tailed distributions such as Pareto \cite{brown_taylors_2021,cohen_covid-19_2022}. Therefore, distributions such as Pareto were widely used for modeling the upper tail of the cases or deaths of the infectious diseases \cite{beare_emergence_2020, blasius_power-law_2020}.   

\par Although Pareto distribution were widely used in modeling the counts or deaths of infectious diseases these years, it can only be used to fit the upper tail of the data since its probability density function is monotonically decreasing. To provide the global fitting of different types of data sets, composite distributions with Pareto tails were utilized. Composite distributions were popularized during the 2000s and became useful in the insurance industry \citep{ananda2005, scollnik_composite_2007,Scollnik2012ModelingWW,brazauskas_modeling_2016, grun2019, liu_analyzing_2022, liu_generalized_2022, mutali_composite_2020, deng_bayesian_2019, ig_pareto, exp_pareto, calderin-ojeda_note_2018}, survival analysis \citep{cooray_weibullpareto_2009}, and reliability engineering \citep{cooray_weibull_2010, liu_new_2023}. Recently, the concept of composite distributions was also utilized to model the cases or deaths due to infectious diseases. For example, the composite lognormal-Pareto distribution was chosen as a proper model, to fit the COVID-19 cases and deaths in the United States \citep{cohen_covid-19_2022}. 

\par However, the cases and deaths of infectious diseases are usually perceived as count data that can only take the the values of non-negative integers. Therefore, the continuous composite distributions that researchers use currently might not be proper when fitting such data. Specifically, if the data contains a considerable number of zeros, such distribution will fail to provide satisfactory performances. 

\par Thus, we intend to create the discretized composite distributions with Pareto tails, using survival discretization method \citep{chakraborty_generating_2015, kemp_classes_2004, roy_discrete_2003}. Our objective is to adapt the original composite distributions to the nature of count data.

\par The rest of the paper will be organized as follows: In Section \ref{Section:2}, we will introduce the method to create discrete composite distributions with Pareto tails. Some special discrete composite distributions with Pareto tails will be presented in Section \ref{Section:3} as well as the tail properties and moment properties of the distributions. Section \ref{Section:4} introduces the parameter estimation and hypothesis testing of the discrete composite distributions. The real data applications to COVID-19 cases and Mpox cases are presented in Section \ref{Section:5}. Conclusions will be given in Section \ref{Section:6}.

\section{Method}
\label{Section:2}
\subsection{Composite Distributions}
The concept of composite distribution was first introduced by Cooray and Ananda\cite{ananda2005}. Later on, composite distribution was formally defined as follows \citep{scollnik_composite_2007}:
\begin{align}
\label{eqn2}
 f_X(x|\boldsymbol{\beta_1},\boldsymbol{\beta_2},a_1, a_2 ,\theta) = \begin{cases} 
      a_1 g_1(x|\boldsymbol{\beta_1},\theta) & \text{if } x \in (0,\theta] \\
      a_2 g_2(x|\boldsymbol{\beta_2},\theta) & \text{if } x \in (\theta,\infty)
   \end{cases},
\end{align}
where $a_1, a_2$ stand for the normalizing constants that guarantees $f_X$ to be a proper pdf, $\theta$ is the single parameter that denotes the location of density change. $g_1$ is the pdf of $X$ if $X$ takes a value that is between $0$ and $\theta$, and $\boldsymbol{\boldsymbol{\beta_1}}$ represents the vector of parameters of $g_1$; $g_2$ is the pdf of $X$ if $X$ takes a value that is greater than $\theta$, and $\boldsymbol{\boldsymbol{\beta_2}}$ represents the vector of parameters of $g_2$. $f_X$ is assumed to be smooth on its support. 
\par Later on, the general composite models were formalized by \cite{Bak15} as follows: \begin{align}
\label{eqn:weighted_comp}
 f_X(x|\boldsymbol{\beta_1},\boldsymbol{\beta_2},\phi,\theta) = \begin{cases} 
      \frac{1}{1+\phi}g_1^*(x|\boldsymbol{\beta_1},\theta) & \text{if } x \in (0,\theta] \\
      \frac{\phi}{1+\phi}g_2^*(x|\boldsymbol{\beta_2},\theta) & \text{if } x \in (\theta,\infty)
   \end{cases},
\end{align}
where $\phi$ is the normalizing constant, $g_1^*$ and $g_2^*$ are truncated pdfs defined as follows:
\begin{align}
\label{eqn3}
\begin{cases}
 g_1^*(x|\boldsymbol{\beta_1},\theta) = \frac{g_1(x|\boldsymbol{\beta_1},\theta)}{G_1(\theta|\boldsymbol{\beta_1},\theta)} \\
  g_2^*(x|\boldsymbol{\beta_2},\theta) = \frac{g_2(x|\boldsymbol{\beta_2},\theta)}{1-G_2(\theta|\boldsymbol{\beta_2},\theta)},
\end{cases}
\end{align}
where $G_1$ and $G_2$ are the corresponding CDF of $g_1$ and $g_2$. 

\par In real practice, we assume that both $g_1$ and $g_2$ are smooth functions on their supports. However, the definition in \ref{eqn3} does not guarantee that $f_X(x)$ is a continuous differentiable function. To define continuous differentiable pdf for composite distributions, continuity and differentiability conditions are taken into account as follows:
\begin{align}
\label{eqn4}
\begin{cases}
\text{lim}_{x \rightarrow \theta^-}f_X(x|\boldsymbol{\beta_1}, \boldsymbol{\beta_2},\phi,\theta)= \text{lim}_{x \rightarrow \theta^+}f_X(x|\boldsymbol{\beta_1}, \boldsymbol{\beta_2},\phi,\theta) \\
\text{lim}_{x \rightarrow \theta^-}\frac{df_X(x|\boldsymbol{\beta_1},\boldsymbol{\beta_2},\phi,\theta)}{dx}= \text{lim}_{x \rightarrow \theta^+}\frac{df_X(x|\boldsymbol{\beta_1},\boldsymbol{\beta_2},\phi,\theta)}{dx}.
\end{cases}
\end{align}
\par Furthermore, with the definitions in \ref{eqn2} and \ref{eqn3}, the conitnuity and differentiability conditions can be written as follows:

                \begin{align}
    \label{eqn5}
\begin{cases}
\phi = \frac{g_1(\theta) [1-G_2(\theta)]}{g_2(\theta)G_1(\theta)} \\
g_1(\theta) g_2^{'}(\theta) = g_2(\theta) g_1^{'}(\theta)
\end{cases}
\end{align}

\par Indeed, the composite distribution can be seen as a special case of spliced distribution with two pieces. The introduction of continuity and differentiability conditions into the model can reduce the total number of unknown parameters and thus reduce the complexity of the spliced distributions. 

\subsection{Problem with differentiability condition with Pareto tails in the composite models}
In real data applications, Pareto distribution is favorable when modeling the tail behaviors due to its ability to capture the heavy tail features of the data. For this reason, the pdf of Pareto distribution was frequently chosen as $g_2$ in Equation \ref{eqn3} and the CDF of Pareto distribution was selected as $G_2$ in Equation \ref{eqn3}:
\begin{align}
\label{eqn6}
\begin{cases}
 g_2(x|\boldsymbol{\beta_2},\theta) = \frac{\alpha \theta^{\alpha}}{x^{\alpha+1}}\\
 G_2(x|\boldsymbol{\beta_2},\theta) = 1-(\frac{\theta}{x})^{\alpha},
\end{cases}
\end{align}
where $\boldsymbol{\beta_2} = \alpha$ in this case. 
Thus, for any $g_1$ and $G_1$ we selected regarding Equation \ref{eqn3}, if the conditions in Equation \ref{eqn4} are enforced, we must have:
\begin{align*}
\begin{cases}
 \phi = \frac{\theta g_1(\theta|\boldsymbol{\beta_1},\theta)}{\alpha G_1(\theta|\boldsymbol{\beta_1},\theta)}\\
 \frac{g_1^{'}(\theta|\boldsymbol{\beta_1},\theta)} {{g_1(\theta|\boldsymbol{\beta_1},\theta)}} = -\frac{\alpha+1}{\theta}
\end{cases}
\end{align*}
Thus, given $\alpha, \theta,$ and $g_1 (\theta|\boldsymbol{\beta_1},\theta)$ are positive, the differentiability condition of the model implies that $g_1^{'}(\theta|\boldsymbol{\beta_1},\theta)<0$. 
\par Notice this is a very strong assumption in pdf $g_1$. For example, suppose that $g_1$ is the pdf of a random variable that follows a uniform distribution in the interval $(0,\theta)$. Then, $g_1(x|\boldsymbol{\beta_1},\theta) = \frac{1}{\theta}$. Therefore, $g_1^{'}(x|\boldsymbol{\beta_1},\theta) = 0$. This essentially tells us that a uniform-Pareto composite distribution cannot be formed with the differentiability condition in Equation \ref{eqn3}. 
\par Thus, for the composite distributions introduced later in Sections \ref{Section:3}, only the continuity condition is used to determine the weight parameter $\phi$. 
\subsection{Survival Discretization Method}

The survival discretization method was widely used to create discrete probability distributions from continuous distributions \cite{chakraborty_generating_2015, kemp_classes_2004, roy_discrete_2003}. In order to create useful discrete models from the common continuous distributions, the method was utilized in the modeling of COVID-19 data recently \cite{almetwally_new_2022, almetwally_overview_2023}. For a continuous random variable $X$ with $[0,\infty)$ support, the probability mass function (pmf) of the corresponding discrete random variable $Y$ can be defined as follows:
\begin{equation}
\label{eqn7}
    P(Y = y) = S_X(y) - S_X(y+1), 
\end{equation}
where $y = 0,1,2,...$ and $S_X$ is the survival function of $X$. Notice $S_X$ must satisfy the following:
\begin{align*}
\begin{cases}
S_X(0) = 1\\
\text{lim}_{x \rightarrow \infty}S_X(x) = 0
\end{cases}
\end{align*}

The pmf of $Y$ is guaranteed to be valid for any continuous random variable $X$ with $[0,\infty)$ support since,
\begin{equation*}
   \sum_{i = 0}^{\infty} P(Y = i) = S_X(0) = 1. 
\end{equation*}
Correspondingly, the cumulative distribution function (CDF) of $Y$ can be derived easily based on the survival function of $X$:
\begin{equation*}
    F_Y(y) = P_Y(Y \leq y) =  \sum_{i = 0}^{y} P(Y = y) = 1- S_X(y+1)
\end{equation*}
Another important property is that the survival function of $Y$ will demonstrate same tail behavior as the survival function of $X$ since for discrete random variable $Y$, the survival function $S_Y$ can be derived as follows:

\begin{equation*}
    S_Y(y) = 1-F_Y(y-1) = 1-[1-S_X(y)] = S_X(y),
\end{equation*}
where $S_X(0) = 1$ since we assume $X$ is a continuous random variable with a support $[0,\infty)$. Hence, $S_Y(0) = 1$ according to the above equation. 

Therefore, since the composite random variables that we discussed in 2.1 and 2.2 are continuous random variables, the survival discretization method can definitely be applied without problems. In the next section, we will introduce some particular composite distributions with Pareto tails and the corresponding discrete composite distributions.

\section{Some Discrete Composite Models with Pareto tails}
\label{Section:3}
In this section, we will introduce two special discrete composite distributions with Pareto tails. Furthermore, we will discuss the moment properties of discrete composite distributions with Pareto tails. 
\subsection{Weighted Discrete Lognormal-Pareto (WDLNP) Model}
Assume $X$ is a random variable following a weighted lognormal-Pareto distribution with continuity condition at $x = \theta$. Then, $g_1$ and $G_1$ in Equation \ref{eqn3} are chosen as: 
  \begin{align*}
  \begin{cases}
    g_1(x|\mu,\sigma)=\frac{1}{\sqrt{2 \pi} \sigma x} e^{-\frac{(\ln x-\mu)^2}  {2 \sigma^2}} \\ 
    G_1(x|\mu, \sigma) = \Phi(\frac{\ln x-\mu}{\sigma})
    \end{cases}
        \end{align*}
Since the distribution of $X$ is associated with a Pareto tail, $g_2$ and $G_2$ are chosen as Equation \ref{eqn5} indicates. 

With the continuity condition, the weight parameter $\phi$ can be expressed as a function of other parameters:
   $$
\phi(\mu, \sigma, \alpha, \theta) = \frac{e^{-\frac{(\ln \theta-\mu)^2}{2 \sigma^2}}}{\sqrt{2 \pi} \alpha \sigma \Phi\left(\frac{\ln \theta-\mu}{\sigma}\right)}
$$ 
Hence, the pdf of $X$ can be written as follows:
  \begin{align*} f_{X}(x|\mu,\sigma,\theta,\alpha) = \begin{cases} 
\frac{1}{1+\phi(\mu, \sigma, \alpha, \theta)}\frac{\frac{1}{\sqrt{2 \pi} \sigma x} e^{-\frac{(\ln x-\mu)^2} {2 \sigma^2}}}{\Phi(\frac{\ln \theta-\mu}{\sigma})}& x \in (0,\theta]\\[10pt]
\frac{\phi(\mu, \sigma, \alpha, \theta)}{1+\phi(\mu, \sigma, \alpha, \theta)}\frac{\alpha \theta^{\alpha}}{x^{\alpha+1}} & x \in (\theta,\infty),\\
\end{cases}
\end{align*}
Using the definition of CDF for a continuous random variable, the survival function of $X$ can be derived as follows:
  \begin{align*}  F_{X}(x|\mu,\sigma,\theta,\alpha)  & = \begin{cases} 
  0 & x = 0 \\[10pt]
\frac{\Phi(\frac{\ln x-\mu}{\sigma})}{[1+\phi(\mu, \sigma, \alpha, \theta)] \Phi(\frac{\ln \theta-\mu}{\sigma})}  &  x \in (0,\theta]\\[10pt]
\frac{1+\phi(\mu, \sigma, \alpha, \theta)[1-(\frac{\theta}{x})^{\alpha}]}{1+\phi(\mu, \sigma, \alpha, \theta)}  &  x \in (\theta,\infty) .\\
\end{cases}
\end{align*}
Correspondingly, the survival function of $X$ is as follows:

\begin{align*}  S_{X}(x|\mu,\sigma,\theta,\alpha)  & = \begin{cases} 
1 & x = 0 \\[10pt]
1-\frac{\Phi(\frac{\ln x-\mu}{\sigma})}{[1+\phi(\mu, \sigma, \alpha, \theta)] \Phi(\frac{\ln \theta-\mu}{\sigma})}   &  x \in (0,\theta]\\[10pt]
1-\frac{1+\phi(\mu, \sigma, \alpha, \theta)[1-(\frac{\theta}{x})^{\alpha}]}{1+\phi(\mu, \sigma, \alpha, \theta)}  &  x \in (\theta,\infty).\\
\end{cases}
\end{align*}

Suppose $Y$ follows a weighted discrete lognormal-Pareto distribution (WDLNP). By utilizing the survival discretization method in \ref{eqn6}, we have the PMF function of the weighted discrete lognormal-Pareto distribution (WDLNP):
\par If $\theta \leq 1$:
 \begin{equation}
    \label{eqn_dwlnp_1}
P(Y = y) = \begin{cases} 
\frac{1+\phi(\mu, \sigma, \alpha, \theta)[1-\theta^{\alpha}]}{1+\phi(\mu, \sigma, \alpha, \theta)}            & y  = 0\\[10pt]

         \frac{\phi(\mu, \sigma, \alpha, \theta)[(\frac{\theta}{y})^{\alpha} -(\frac{\theta}{y+1})^{\alpha}]}{1+\phi(\mu, \sigma, \alpha, \theta)} & y \in (0,\infty).
       \end{cases} 
  \end{equation}
  
If $\theta > 1$: 
 \begin{equation}
    \label{eqn_dwlnp_2}
P(Y = y) = \begin{cases} 
          \frac{\Phi(\frac{\ln (y+1)-\mu}{\sigma})-\Phi(\frac{\ln y-\mu}{\sigma})}{[1+\phi(\mu, \sigma, \alpha, \theta)] \Phi(\frac{\ln \theta-\mu}{\sigma})}& y \in [0, \theta-1) \\[10pt]
          \frac{ \Phi(\frac{\ln \theta-\mu}{\sigma})+\phi(\mu, \sigma, \alpha, \theta)\Phi(\frac{\ln \theta-\mu}{\sigma})[1-(\frac{\theta}{y+1})^{\alpha}]-\Phi(\frac{\ln y-\mu}{\sigma})}{[1+\phi(\mu, \sigma, \alpha, \theta)] \Phi(\frac{\ln \theta-\mu}{\sigma})}  &y \in [\theta-1,\theta)\\[10pt]

         \frac{\phi(\mu, \sigma, \alpha, \theta)[(\frac{\theta}{y})^{\alpha} -(\frac{\theta}{y+1})^{\alpha}]}{1+\phi(\mu, \sigma, \alpha, \theta)} & y \in [\theta,\infty). 
       \end{cases} 
  \end{equation}

where $y$ takes values on nonnegative integers. Notice that when $y=0$ and $\theta > 1$, $P(Y=0) = \text{lim}_{y \rightarrow 0}  \frac{\Phi(\frac{\ln (y+1)-\mu}{\sigma})-\Phi(\frac{\ln y-\mu}{\sigma})}{[1+\phi(\mu, \sigma, \alpha, \theta)] \Phi(\frac{\ln \theta-\mu}{\sigma})} = \frac{\Phi(-\frac{\mu}{\sigma})}{[1+\phi(\mu, \sigma, \alpha, \theta)] \Phi(\frac{\ln \theta-\mu}{\sigma})}$ since $\text{lim}_{y \rightarrow 0} \Phi(\frac{\ln y-\mu}{\sigma}) = 0$.

\par Essentially, $Y$ shows a very similar behavior to $X$ since $S_Y(y) = S_X(y)$ for $y = 0, 1, 2, ...$. Since $X$ follows a heavy-tailed distribution due to the Pareto tail, $Y$ automatically follows a discrete heavy-tailed distribution. In addition, $Y$ follows a discrete distribution with a fat tail when $\alpha<2$. 

    \begin{figure}[htbp]
\includegraphics[width=12cm]{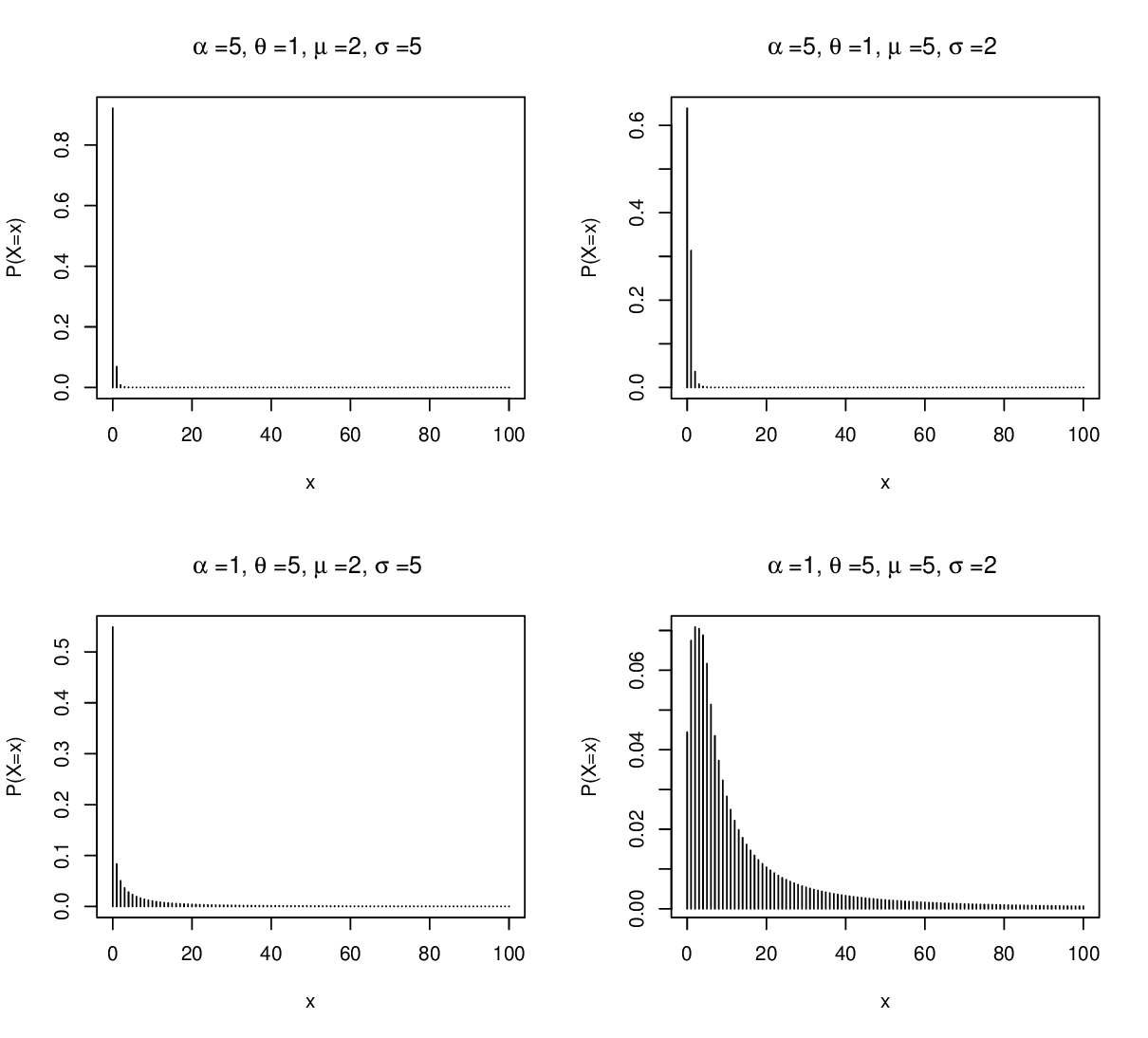}
\centering
\caption{The PMF plot of DWLNP with different parameters} %four parameter settings
\end{figure}

\subsection{Weighted Discrete Weibull-Pareto (WDWP) Model}

Suppose $X$ is a random variable that follows a weighted Weibull-Pareto (WWP) distribution that utilizes only the continuity condition in $x = \theta$. Then in Equation \ref{eqn3}, we choose the pdf and CDF of Weibull distribution as $g_1$ and $G_1$.

\begin{align*}
\begin{cases}
g_1(x)= (\frac{a}{\sigma})(\frac{x}{\sigma})^{a-1} e^ {-(\frac{x}{\sigma})^{a}} \\[10pt]
G_1(x) = 1-e^{-(\frac{x}{\sigma})^a}
\end{cases}
\end{align*}

We also utilize Equation \ref{eqn5} as our selection for $g_2$ and $G_2$. 
Similar to $\phi$ parameter in the WWP model, a closed-form expression of $\phi$ in terms of all other parameters:
\begin{equation*}  
 \phi(a,\sigma,\alpha,\theta)=\frac{a \theta^a e^{-\left(\frac{\theta}{\sigma}\right)^a}}{\alpha \sigma^a\left[1-e^{-\left(\frac{\theta}{\sigma}\right)^a}\right].}
\end{equation*}

The pdf of WWP can then be expressed as follows with 4 parameters: 

\begin{align*}f_{X}(x|a,\sigma,\alpha,\theta) = \begin{cases} 
\frac{1}{1+\phi(a,\sigma,\alpha,\theta)}\frac{ (\frac{a}{\sigma})(\frac{x}{\sigma})^{a-1} e^ {-(\frac{x}{\sigma})^{a}}}{1-e^{-(\frac{\theta}{\sigma})^a}}& x \in (0,\theta]\\[10pt]
\frac{\phi(a,\sigma,\alpha,\theta)}{1+\phi(a,\sigma,\alpha,\theta)}\frac{\alpha \theta^{\alpha}}{x^{\alpha+1}} & x \in (\theta,\infty),\\
\end{cases}
\end{align*}

The CDF of WWP can be derived as follows based on its definition:
\begin{align*}F_{X}(x|a,\sigma,\alpha,\theta) = \begin{cases} 
0 & x = 0 \\[10pt]
\frac{1}{1+\phi(a,\sigma,\alpha,\theta)}\frac{1-e^{-(\frac{x}{\sigma})^a}}{1-e^{-(\frac{\theta}{\sigma})^a}}& x \in (0,\theta]\\[10pt]
\frac{1+\phi(a,\sigma,\alpha,\theta)[1-(\frac{\theta}{x})^{\alpha}]}{1+\phi(a,\sigma,\alpha,\theta)}  &  x \in (\theta,\infty) .\\
\end{cases}
\end{align*}

The survival function of WWP can then be derived as:
\begin{align*} S_{X}(x|a,\sigma,\alpha,\theta) = \begin{cases} 
1 & x = 0 \\[10pt]
1-\frac{1}{1+\phi(a,\sigma,\alpha,\theta)}\frac{1-e^{-(\frac{x}{\sigma})^a}}{1-e^{-(\frac{\theta}{\sigma})^a}}& x \in (0,\theta]\\[10pt]
1-\frac{1+\phi(a,\sigma,\alpha,\theta)[1-(\frac{\theta}{x})^{\alpha}]}{1+\phi(a,\sigma,\alpha,\theta)}  &  x \in (\theta,\infty) .\\
\end{cases}
\end{align*}

Assume $Y$ is a random variable that follows WDWP distribution. Using the survival discretization method, the corresponding pmf of the WDWP model can be obtained as follows:

\par If $\theta \leq 1$:
 \begin{equation}
    \label{eqn_wdwp_1}
P(Y = y) = \begin{cases} 
\frac{1+\phi(a,\sigma,\alpha,\theta)[1-\theta^{\alpha}]}{1+\phi(a,\sigma,\alpha,\theta)}            & y  = 0\\[10pt]

         \frac{\phi(a,\sigma,\alpha,\theta)[(\frac{\theta}{y})^{\alpha} -(\frac{\theta}{y+1})^{\alpha}]}{1+\phi(a,\sigma,\alpha,\theta)} & y \in (0,\infty) ,
       \end{cases} 
  \end{equation}
\par If $\theta > 1$, 
\begin{equation}
    \label{eqn_wdwp_2}
P(Y = y) = \begin{cases} 
          \frac{e^{-(\frac{y}{\sigma})^a}-e^{-(\frac{y+1}{\sigma})^a}}{[1+\phi(a,\sigma,\alpha,\theta)] [1-e^{-(\frac{\theta}{\sigma})^a}]}& y \in [0, \theta-1) \\[10pt]

           \frac{e^{-(\frac{y}{\sigma})^a}-e^{-(\frac{\theta}{\sigma})^a}+\phi(a,\sigma,\alpha,\theta)[1-e^{-(\frac{\theta}{\sigma})^a}][1-(\frac{\theta}{y+1})^\alpha]}{[1+\phi(a,\sigma,\alpha,\theta)] [1-e^{-(\frac{\theta}{\sigma})^a}]}  &y \in [\theta-1,\theta)\\[10pt]

         \frac{\phi(a,\sigma,\alpha,\theta)[(\frac{\theta}{y})^{\alpha} -(\frac{\theta}{y+1})^{\alpha}]}{1+\phi(a,\sigma,\alpha,\theta)}& y \in [\theta,\infty). 
       \end{cases} 
  \end{equation}
\par  Similar to WDLNP model, $Y$ follows a discrete distribution with a heavy tail. The distribution of $Y$ has a fat tail when $\alpha< 2$.
    \begin{figure}[htbp!]
\includegraphics[width=12cm]{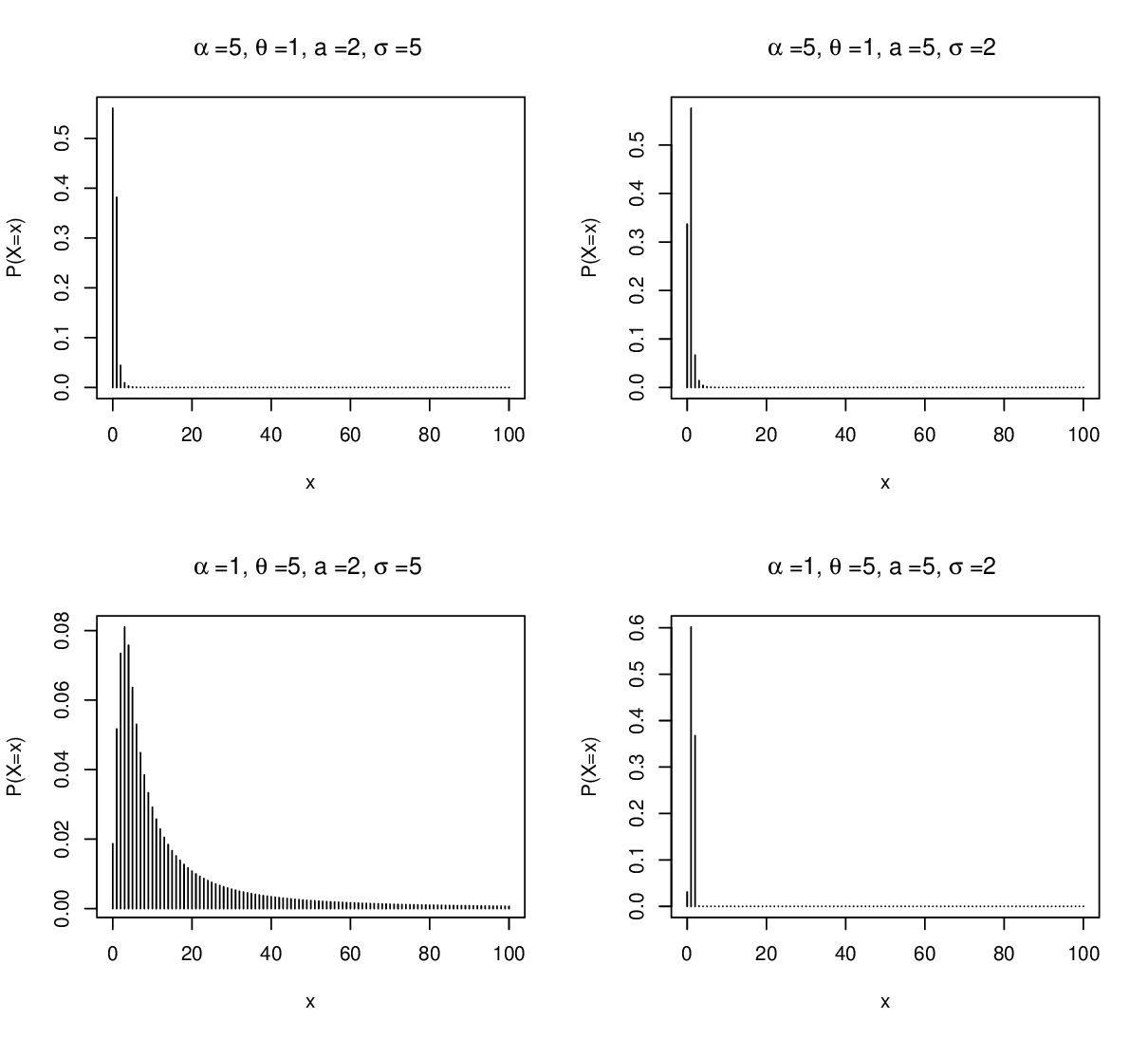}
\centering

\caption{The PMF plot of DWWP with different parameters} %four parameter settings
\end{figure}

\subsection{The properties of the $n$-th moments for the discrete composite distributions with Pareto tails}
\label{subsec: moment}
In this subsection, important properties of the discrete composite distributions with Pareto tails are discussed in terms of $n$-th moments. As an analogue of continuous composite distributions with Pareto tails, we intend to formally show that the moments of discrete composite distributions with Pareto tails only exist under certain conditions. 

\begin{theorem}\label{n-th moment}
    Suppose $Y$ is associated with a weighted discrete composite distribution with a Pareto tail, then the $n$-th moment of $Y$ does not exist if $\alpha \leq n$, where $\alpha$ is the shape parameter of the Pareto tail. 
\end{theorem}

\begin{pf}
    Suppose $Y$ follows a weighted discrete composite distribution created from Equation \ref{eqn:weighted_comp} with only the continuity condition, using the survival disretization method in \ref{eqn6}. 
    Then, the $n$-th moment of $Y$ can be written as follows:
    \begin{align*}
    \begin{split}
    E(Y^n) & = \sum_{y \in \mathbb{N}_0, y< \theta} y^n P(Y=y|Y < \theta) + \sum_{y \in \mathbb{N}_0, y \geq \theta} y^n P(Y=y|Y \geq \theta) \\
    & \geq \sum_{y \in \mathbb{N}_0, y \geq \theta} y^n P(Y=y|Y \geq \theta),
     \end{split}
    \end{align*}
    where $\theta$ is the parameter that determines the change of density in a composite distribution. 
    
    Notice for any discrete composite random variable $Y$ with a Pareto($\alpha, \theta$) tail that created based on a continuous composite distribution with the survival dicretization method, we have the following: 
    $$P(Y = y| Y \geq \theta) = \frac{\phi[(\frac{\theta}{y})^{\alpha} -(\frac{\theta}{y+1})^{\alpha}]}{1+\phi},$$
    where $\phi$ is uniquely determined by other parameters of the model. 

    Then, 
     \begin{align*}
    \begin{split}
    \sum_{y \in \mathbb{N}_0, y \geq \theta} y^n P(Y=y|Y \geq \theta) & = \sum_{y \in \mathbb{N}_0, y \geq \theta} y^n \frac{\phi[(\frac{\theta}{y})^{\alpha} -(\frac{\theta}{y+1})^{\alpha}]}{1+\phi}\\
    & = \frac{\phi \theta^\alpha}{1+\phi}\sum_{y \in \mathbb{N}_0, y \geq \theta} \left[ y^{n-\alpha}- y^{n}(y+1)^{-\alpha} \right]\\
    & \geq \frac{\phi \theta^\alpha}{1+\phi}\sum_{y \in \mathbb{N}_0, y \geq \theta} \left[ y^{n-\alpha}- (y+1)^{n-\alpha} \right]\\
    & = \frac{\phi \theta^\alpha}{1+\phi}\left[ {\lceil \theta \rceil}^{n-\alpha} - \lim_{y \rightarrow \infty} (y+1)^{n-\alpha} \right],
     \end{split}
    \end{align*}
where $\lceil \theta \rceil$ is the least integer greater or equal to $\theta$. Notice $\lim_{y \rightarrow \infty} (y+1)^{n-\alpha}$ is infinite if $n-\alpha \geq 0 $. Therefore, the $n$-th moment does not exist for $Y$ if $\alpha \leq n$. 
\end{pf}
As a consequence, we have the following two corollaries:
\begin{cor}
Suppose $Y$ is associated with a weighted discrete composite distribution with a Pareto tail, the mean of $Y$ does not exist if $\alpha \leq 1$.
\end{cor}

\begin{cor}
Suppose $Y$ is associated with a weighted discrete composite distribution with a Pareto tail, the variance of $Y$ does not exist if $\alpha \leq 2$.
\end{cor}

In fact, for some special choices of parameters, we are able to derive the moment explicitly. For example, when $\alpha>1$ and $\theta \leq 1$, we can derive the following:
\begin{theorem}
Suppose $Y$ follows a weighted discrete composite distribution with a Pareto tail. Assume $\alpha> 1$ and $\theta \leq 1$. The mean of $Y$ is equal to $\frac{\phi \theta^\alpha}{1+\phi} \zeta(\alpha)$, where $\zeta(.)$ stands for a Riemann zeta function. 
\end{theorem}
\begin{pf}
Suppose $\theta \leq 1$ and $\alpha>1$, then we have the following:
    \begin{align*}
    \begin{split}
    E(Y) & = \sum_{y \in \mathbb{N}_0, y< \theta} y P(Y=y|Y < \theta) + \sum_{y \in \mathbb{N}_0, y \geq \theta} y^n P(Y=y|Y \geq \theta) \\
    & =\sum_{y^n \in \mathbb{N}_0, y \geq \theta} yP(Y=y|Y \geq \theta) \\ 
    & = \sum_{y=1}^{\infty} yP(Y=y|Y \geq \theta)\\ 
    & = \sum_{y=1}^{\infty} y \frac{\phi[(\frac{\theta}{y})^{\alpha} -(\frac{\theta}{y+1})^{\alpha}]}{1+\phi} \\
    & = \frac{\phi \theta^\alpha}{1+\phi} \sum_{y=1}^{\infty} \left[ y^{1-\alpha}- y(y+1)^{-\alpha} \right] \\
    & = \frac{\phi \theta^\alpha}{1+\phi} \left[ 1+\sum_{y=2}^{\infty} y^{1-\alpha} -\sum_{y=2}^{\infty} (y-1)y^{-\alpha}  \right]\\
    & = \frac{\phi \theta^\alpha}{1+\phi} \left(1+ \sum_{y=2}^{\infty} \frac{1}{y^{\alpha}} \right) \\
    & = \frac{\phi \theta^\alpha}{1+\phi}\left(\sum_{y=1}^{\infty} \frac{1}{y^{\alpha}} \right) \\
    & = \frac{\phi \theta^\alpha}{1+\phi} \zeta(\alpha)
     \end{split}
    \end{align*}
\end{pf}

Since $\zeta(2) = \frac{\pi^2}{6}$, we have the following: 

\begin{rm}
Suppose $Y$ follows a weighted discrete composite distribution with a Pareto tail. Suppose $\theta \leq 1$ and $\alpha = 2$. The mean of $Y$ is equal to $\frac{ \pi^2 \phi \theta^2 }{6(1+\phi)}$
\end{rm}

In the next section, the inference on the scale parameter $\alpha$ will be introduced based on properties we discussed in this section. 

\section{Estimation and Hypothesis Testing}
In this section, the parameter estimation of the discrete composite distributions and the hypothesis testing procedures of the tail index will be introduced.
\label{Section:4}
\subsection{Estimation}
Given a particular data set, the parameters the discrete composite distributions with Pareto tails can be estimated with maximum likelihood estimation. Suppose $y_1, y_2, ..., y_n$ is a random sample of size $n$ that follows a weighted discrete composite distribution with a Pareto tail. For different $\theta$ values, the likelihood can be written as follows:

\begin{enumerate}
    \item When $\theta \leq 1$, the pmf of $y_i$ can be written as follows:
    
        \begin{equation*}
P(Y = y_i) = \begin{cases} 
\frac{1+\phi(\boldsymbol{\beta_1},\alpha,\theta)[1-\theta^{\alpha}]}{1+\phi(\boldsymbol{\beta_1},\alpha,\theta)}            & y_i  = 0\\[10pt]

         \frac{\phi(\boldsymbol{\beta_1},\alpha,\theta)[(\frac{\theta}{y_i})^{\alpha} -(\frac{\theta}{y_i+1})^{\alpha}]}{1+\phi(\boldsymbol{\beta_1},\alpha,\theta)} & y_i \in (0,\infty) ,
       \end{cases} 
  \end{equation*}
  where $\alpha$ and $\theta$ are the parameters of the Pareto tail; $\boldsymbol{\beta_1}$ is the parameter vector for the distribution selected as the head in a weighted discrete composite distribution. 
Then, the corresponding log-likelihood $l_1$ can be expressed as follows:
\begin{equation*}
\begin{aligned}
l_1 & = -n \text{ln}[1+\phi(\boldsymbol{\beta_1},\alpha,\theta)]+\sum_{i=1}^{m} \text{ln}\left\{1+\phi(\boldsymbol{\beta_1},\alpha,\theta)(1-\theta^\alpha)\right \}\\
&+\sum_{y_i >0} \text{ln} \left\{\phi(\boldsymbol{\beta_1},\alpha,\theta)[(\frac{\theta}{y_i})^{\alpha} -(\frac{\theta}{y_i+1})^{\alpha}]\right\},
\end{aligned}
  \end{equation*}
  where $m = \sum_{i=1}^{n}I(y_i=0)$ and $I(.)$ stands for the indicator function. 
    \item When $\theta > 1$, the pmf of $y_i$ can be represented as follows: 

    \begin{equation*}
P(Y = y) = \begin{cases} 
    \frac{G_1(y+1|\boldsymbol{\beta_1}) - G_1(y|\boldsymbol{\beta_1})}{[1+\phi(\boldsymbol{\beta_1},\alpha,\theta)] [G_1(\theta|\boldsymbol{\beta_1})]}& y \in [0, \theta-1) \\[10pt]

           \frac{G_1(\theta|\boldsymbol{\beta_1})-G_1(y|\boldsymbol{\beta_1})+\phi(\boldsymbol{\beta_1},\alpha,\theta)[G_1(\theta|\boldsymbol{\beta_1})][1-(\frac{\theta}{y+1})^\alpha]}{[1+\phi(\boldsymbol{\beta_1},\alpha,\theta)] [G_1(\theta|\boldsymbol{\beta_1})]}  &y \in [\theta-1,\theta)\\[10pt]

         \frac{\phi(\boldsymbol{\beta_1},\alpha,\theta)[(\frac{\theta}{y})^{\alpha} -(\frac{\theta}{y+1})^{\alpha}]}{1+\phi(\boldsymbol{\beta_1},\alpha,\theta)}& y \in [\theta,\infty),
       \end{cases} 
  \end{equation*}
    where $G_1$ is the cdf of the distribution selected as the head. 

Thus, the corresponding log-likelihood $l_2$ can be represented as follows:
\begin{equation*}
\begin{aligned}
l_2 & = -n \text{ln}[1+\phi(\boldsymbol{\beta_1},\alpha,\theta)]+\sum_{y_i < \theta-1} \text{ln}\left[ G_1(y_i+1|\boldsymbol{\beta_1}) - G_1(y_i|\boldsymbol{\beta_1})\right ]\\
&+ \sum_{\theta-1 \leq y_i < \theta}\text{ln} \left\{ G_1(\theta|\boldsymbol{\beta_1})-G_1(y_i|\boldsymbol{\beta_1})+\phi(\boldsymbol{\beta_1},\alpha,\theta)[G_1(\theta|\boldsymbol{\beta_1})][1-(\frac{\theta}{y_i+1})^\alpha] \right\} \\
&- (m_1+m_2)\text{ln}[G_1(\theta|\boldsymbol{\beta_1})] +\sum_{y_i \geq \theta} \text{ln} \left\{\phi(\boldsymbol{\beta_1},\alpha,\theta)[(\frac{\theta}{y_i})^{\alpha} -(\frac{\theta}{y_{i}+1})^{\alpha}]\right\},
\end{aligned}
  \end{equation*}
    
\end{enumerate}
where $m_1 = \sum_{i=1}^{n}I(y_i < \theta-1)$ and $m_2 = \sum_{i=1}^{n} I(\theta-1 \leq y_i < \theta)$.

Eventually, the log-likelihood of $y_1, y_2, ..., y_n$ can be represented as follows with $l_1$ and $l_2$ we just derived:

\begin{equation}
\label{loglik}
l(\boldsymbol{\beta_1},\alpha,\theta|y_1, y_2, ..., y_n) = I(\theta \leq 1)l_1+[1-I(\theta\leq 1)]l_2. 
\end{equation}

The maximum likelihood estimates (MLE) of the parameters are as follows:

\begin{equation*}
 (\hat{\boldsymbol{\beta_1}},\hat{\alpha},\hat{\theta}) =  \argmax_{\boldsymbol{\beta_1},\alpha,\theta} l(\boldsymbol{\beta_1},\alpha,\theta|y_1, y_2, ..., y_n) 
\end{equation*}

Clearly, the MLE of the parameters cannot be obtained with an analytical form. With the 'optim' function in R, we can obtain the MLE numerically. 

\subsection{Testing of the shape parameter $\alpha$}
\label{sec:4.2}
In Section \ref{Section:3}, it is shown that the moment a discrete composite random variable with a Pareto tail does not exist when $\alpha \leq 1$. We also show that the variance of a discrete composite random variable with a Pareto tail does not exist when $\alpha \leq 2$. Numerically, we are able to find the MLE of $\alpha$ by maximizing the loglikelihood in Equation \ref{loglik}. In addition to the sole point estimate of $\alpha$, hypothesis testing of the shape parameter $\alpha$ can be quite helpful in addition to the sole point estimate of $\alpha$ in real data applications to the infectious disease counting data. Essentially, the null hypotheses of interest are as follows: 

\begin{enumerate}

    \item $H_0: \alpha \leq 1$ (Test for infinite mean)
    \item $H_0: \alpha \leq 2$ (Test for infinite variance)
\end{enumerate}

\par In order to test the above hypotheses, the likelihood ratio test (LRT) can be used. Assume that the loglikelihood function $l(\boldsymbol{\beta_1},\alpha,\theta|y_1, y_2, ..., y_n)$ takes the form in Equation \ref{loglik}. For the simplicity of the expression, let $\Pi = (\boldsymbol{\beta_1},\alpha,\theta)$ and $\boldsymbol{y} = (y_1, y_2, ..., y_n)$. Let $S$ be the parameter space of $\Pi$ and $S_0$ be the parameter space of $\Pi$ when $H_0$ is true. The LRT statistic $\lambda$ for the given hypothesis is as follows:
\begin{equation}\label{eqn:3.17}
\centering
    \lambda = 2 \text{sup}_{\Pi \in S}l(\Pi|\boldsymbol{y})- 2 {\text{sup}_{\Pi \in S_0}l(\Pi|\boldsymbol{y})} ,
\end{equation}
where $\lambda$ follows a $\chi^2$ distribution with a degree of freedom 1 asymptotically. The corresponding p-value of the test is $P(\chi^2_1> \lambda)$. $H_0$ shall be rejected when the p-value is less than the prespecified significance level. 

\section{Applications}
\label{Setion:5}
In this section, the two novel discrete composite distributions are applied to two different infectious disease data sets. For comparison purposes, we also analyze the data sets with four well-known parametric models: Poisson, zero-inflated Poisson (ZIP), negative binomial (NB), and zero-inflated negative binomial (ZINB). To evaluate model performance, we use the Akaike Information Criteria (AIC) \cite{burnham}. AIC is defined as follows :
\begin{center}
	$AIC = -2l+2k$,
\end{center}
where $l$ denotes the loglikelihood. $k$ denotes the number of free parameters. A model with the lowest AIC value is considered a preferred fit when compared to all the chosen models.  
\par In addition, we also utilize the hypothesis testing procedures in \ref{sec:4.2} to provide statistical inference of the tail index parameter for the two discrete composite distributions with Pareto tails. 
\label{Section:5}

\subsection*{Case 1: COVID-19 Outbreak: Singapore}

\subsubsection*{Data Description and Analysis Plan}
\begin{figure}[pos=htbp!]
    
\includegraphics[width=12cm]{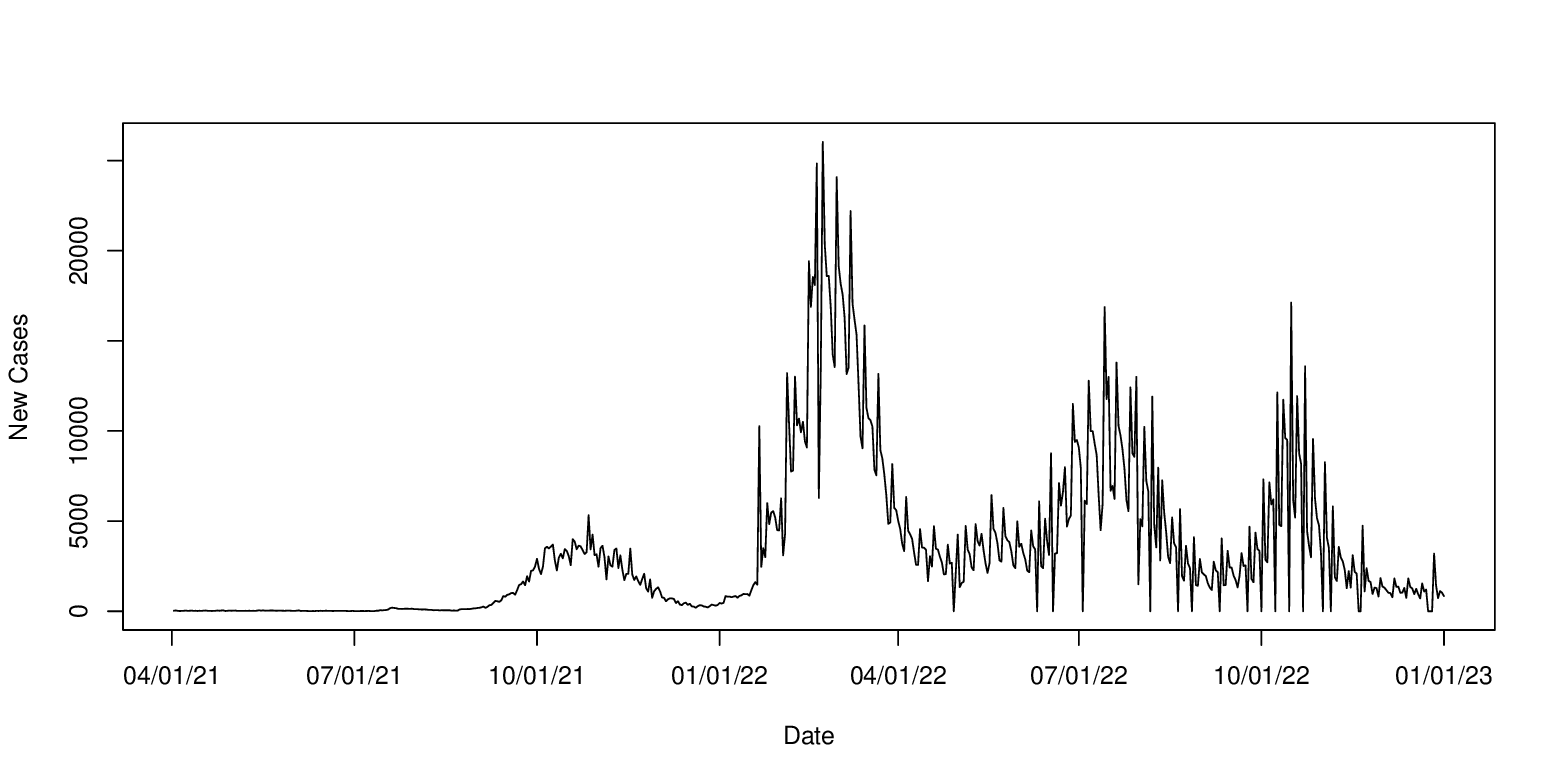}
\centering
\caption{The COVID-19 cases over time in Singapore from 04/2021 to 01/2023}
\label{fig:cs_sin}
\end{figure}

To evaluate the performance of the new discrete composite distributions, the COVID-19 data set on  WHO website was utilized. The data set is publicly available at \url{https://covid19.who.int/data}. The numbers of reported cases by days in Singapore from April 1st, 2021 to January 1st, 2023 were included in our analysis. The time series for the daily count of cases is presented in Figure \ref{fig:cs_sin}. We first analyze the reported cases from the first three months (4/1/2021-7/1/2021). Then, we update our analysis by incorporating the cases from the next three months. In a sequential manner, we were able to obtain the analysis results on seven particular dates: 7/1/2021, 10/1/2021, 1/1/2022, 4/1/2022, 7/1/2022, 10/1/2022, and 1/1/2023. 

\subsubsection*{Results}
\begin{figure}[pos = H]
    \includegraphics[width=12cm]{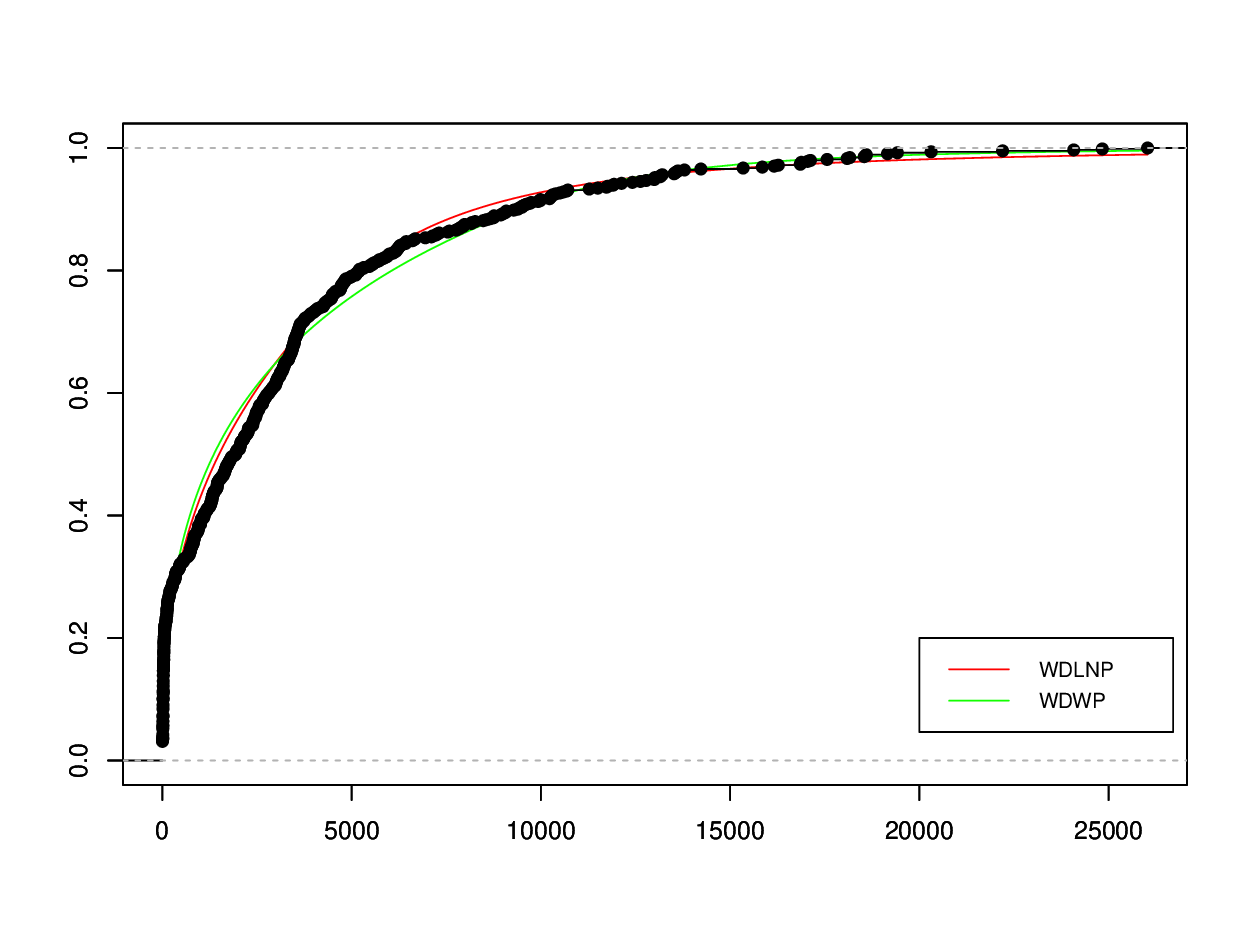}
\centering
\caption{The empirical CDF of COVID-19 cases in Singapore from 4/1/2021 to 1/1/2023 with the fitted CDF of two discrete composite distributions}
\label{fig:4}
\end{figure}

Table \ref{table:1} shows the AIC values for six different models on seven particular dates. Models with the lowest AIC values were highlighted in red. For the dates 10/1/2021, 04/1/2022, 10/1/2022, and 1/1/2023, we noticed that new discrete composite models (WDLNP and WDWP) can provide better performance compared to other well-known models. Figure \ref{fig:4} shows the estimated CDF plot with two discrete composite models with the empirical CDF when we included all daily reported count cases from 4/1/2021 to 1/1/2023. Both models demonstrated great ability in fitting the data, especially at the upper tail. 

\begin{figure}[pos = H]

\includegraphics[width=12cm]{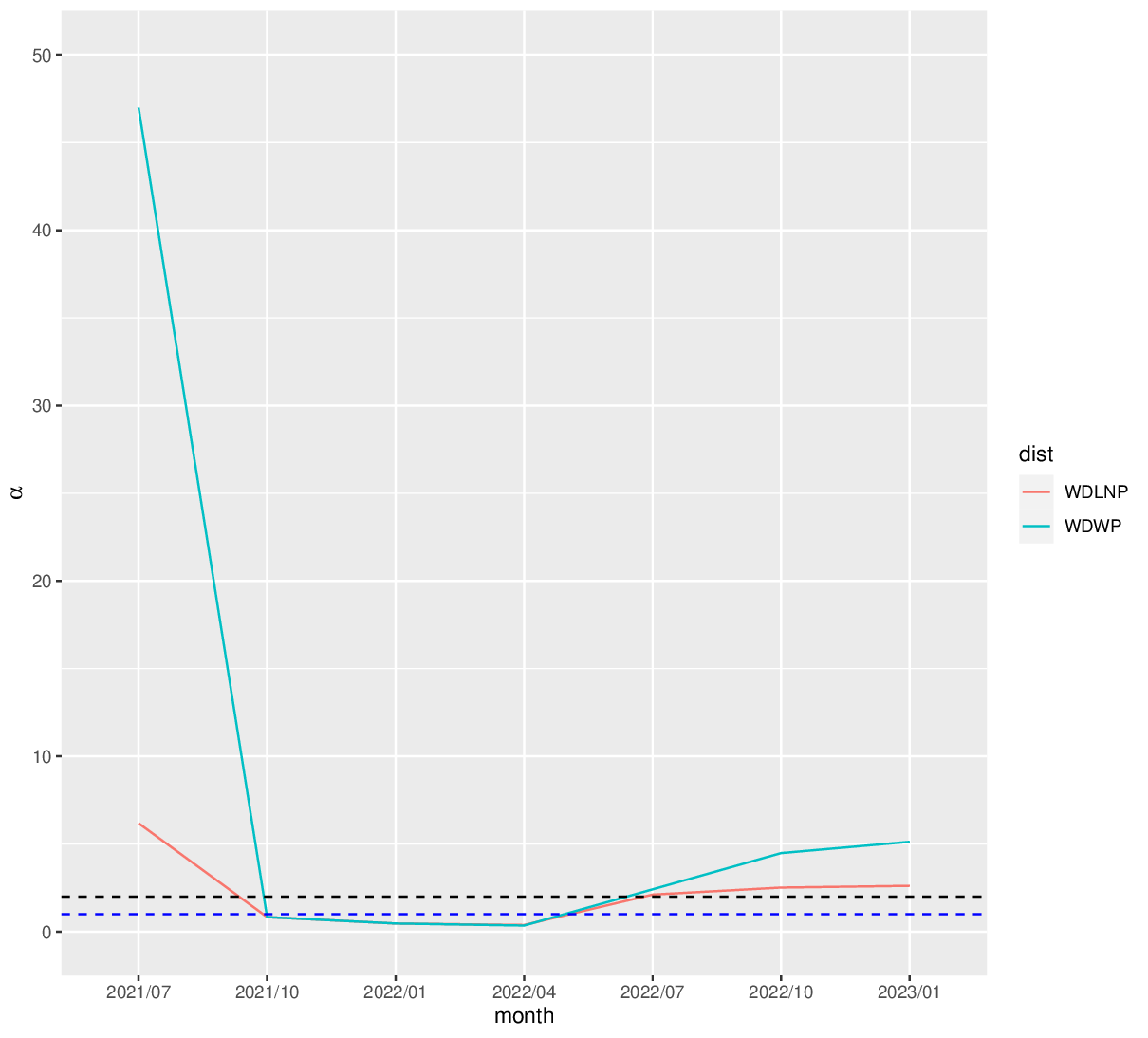}
\centering
\caption{The change of $\alpha$ estimates over time for two composite models when fitting to the COVID-19 cases data from Singapore}
\label{fig:5}
\end{figure}

Figure \ref{fig:5} demonstrates the change of $\alpha$ estimates over time for two discrete composite models. The black horizontal dashed line stands for $\alpha = 2$, and the blue horizontal dashed line represents $\alpha = 1$. Note for both models, the $\alpha$ estimates are less than $1$ during 10/1/2021 and 04/01/2022. Table \ref{table:2} and \ref{table:3} shows the results of the hypothesis testing for the tail index parameter $\alpha$ for two models. During 10/1/2021 and 04/01/2022, we notice that the null hypothesis of $\alpha=1$ and $\alpha=2$ should not be rejected for both models at the significance level of 0.05. This suggests that during this period of time, the estimated mean and variance of the case counts for both models are likely to be infinite. On 7/1/2022, only $\alpha = 1$ should be rejected for both models. This implies that the mean of the case counts is now finite, but still likely to be associated with infinite variance. On 10/1/2022 and 1/1/2023, both null hypothesis for the tail index 
 should be rejected for the WDWP model. The $p$-value of testing $\alpha = 2$ for the WDLNP model is inconclusive on 10/1/2022. However, both null hypothesis should be rejected at the level of 0.05 for the WDLNP model. 

 If we examine the results from Figure \ref{fig:5}, Table \ref{table:2}, and Table \ref{table:3} together with the time series plot in \ref{fig:cs_sin}, we are able to get a general understanding about the severity of pandemic in Singapore in terms of case counts. The COVID-19 pandemic became a major public health problem in Singapore during 10/1/2021 and 4/1/2022, as the tail index estimates $\alpha$ are below 1. On 7/1/2022, although the estimate of $\alpha$ is beyond 2 for both composite models, we cannot still reject $\alpha \leq 2$ based on the LRT test. This implies it is still very likely to see rare but extremely large number of counts of cases on a single day. The local hospitals and public health agencies should still be alert to very large counts. From 10/1/2022, although we still see a considerable amount of reported cases in Figure \ref{fig:cs_sin}, we have gained more evidence that $\alpha \leq 2$ will be rejected for both models. Essentially, this implies that the local health agencies were likely to be prepared for these new cases, since they have already survived from a higher peak of reported cases. 
\begin{table}[H]
\caption{AIC values for six different models (COVID cases in Singapore: 4/1/2021-1/1/2023)}

\label{table:1}
\begin{tabular}{lllllll}
\hline
                 & \textbf{Poisson} & \textbf{ZIP}\textsuperscript{2} & \textbf{NB}\textsuperscript{3}                     & \textbf{ZINB}\textsuperscript{4} & \textbf{WDLNP}\textsuperscript{5}                  & \textbf{WDWP}\textsuperscript{6}                   \\ \hline
\textbf{Date}\textsuperscript{1}    &                  &               &                                 &                &              &                                                               \\ \hline
\textit{7/1/2021}  & 846.8136         & 848.8143      & {\color[HTML]{FE0000} 683.1605} & 685.1605           & 691.7478                        & 689.6392                        \\ \hline
\textit{10/1/2021} & 101068.4         & 101070.4      & {\color[HTML]{333333} 2252.341} & 2254.341          & {\color[HTML]{FE0000} 2096.286} & 2096.432                        \\ \hline
\textit{1/1/2022}  & 416336.8         & 416338.8      & 4027.38                         & 4029.38             & {\color[HTML]{333333} 3967.172} & {\color[HTML]{FE0000} 3967.144} \\ \hline
\textit{4/1/2022}  & 2249692          & 2249694       & 6092.806                        & 6094.806            & {\color[HTML]{FE0000} 6050.2}   & 6079.752                        \\ \hline
\textit{7/1/2022}  & 2366642          & 2366644       & {\color[HTML]{FE0000} 7850.405} & 7852.405             & 7856.288                        & 7854.078                        \\ \hline
\textit{10/1/2022} & 2706326          & 2706328       & 9610.242                        & 9612.242            & {\color[HTML]{FE0000} 9603.92}  & 9605.378                        \\ \hline
\textit{1/1/2023} & 2995122          & 2859516       & 11245.88                        & 11244.79            & {\color[HTML]{FE0000} 11231.17} & 11239.7                         \\ \hline
\end{tabular}

\noindent{ \footnotesize{\textsuperscript{1} The data was included from 4/1/2021 until this date.}}\\
	\noindent{ \footnotesize{\textsuperscript{2} Zero-inflated Poisson}}\\
 \noindent{ \footnotesize{\textsuperscript{3} Negative binomial}}\\
  \noindent{ \footnotesize{\textsuperscript{4} Zero-inflated negative binomial}}\\

  \noindent{ \footnotesize{\textsuperscript{5} Weighted discrete lognormal-Pareto}}\\
  \noindent{ \footnotesize{\textsuperscript{6} Weighted discrete Weibull-Pareto}}
\end{table}

%alpha test tables: WDLNP Singapore
\begin{table}[pos=H]
\caption{The LRT test results for tail index $\alpha$ with WDLNP model when fitting to COVID-19 cases in Singapore}
\label{table:2}
\begin{tabular}{@{}lllll@{}}
\toprule
\textbf{Date}      & \multicolumn{2}{c}{$\boldsymbol{H_0: \alpha \leq 1}$}                                                 & \multicolumn{2}{c}{$\boldsymbol{H_0: \alpha \leq 2}$}                                                          \\ \midrule
\textbf{}          & \multicolumn{1}{c}{$\chi^2$\textit{\textbf{ Statistic}}} & \multicolumn{1}{c}{\textit{\textbf{p-value}}} & \multicolumn{1}{c}{$\chi^2$\textit{\textbf{ Statistic}}} & \multicolumn{1}{c}{\textit{\textbf{p-value}}} \\ \cmidrule(l){2-5} 
\textit{7/1/2021}  & 77.52                                              & $1.31 \times 10^{-18}$                           & 28.79                                              & $8.05 \times 10^{-8}   $                                  \\
\textit{10/1/2021} & 0                                                  & 1                                    & 0                                                  & 1                                             \\
\textit{1/1/2022}  & 0                                                  & 1                            & 0                                                  & 1                                     \\
\textit{4/1/2022}  & $4.85 \times 10^{-8}$                                          & 1                                    & 0                                                  & 1                                             \\
\textit{7/1/2022}  & 48.81                                              & $2.82 \times 10^{-12}$                             & 0.37                                               & 0.54                                          \\
\textit{10/1/2022} & 70.38                                              & $4.89 \times 10^{-17} $                            & 3.49                                               & 0.06                                          \\
\textit{1/1/2023}  & 85.59                                              & $2.22 \times 10^{-20}  $                           & 5.98                                               & 0.01                                          \\ \bottomrule
\end{tabular}
\end{table}  

%alpha test tables: WDWP Singapore
    \begin{table}[pos=H]
    \caption{The LRT test results for tail index $\alpha$ with WDWP model when fitting to COVID-19 cases in Singapore}
    \label{table:3}
\begin{tabular}{@{}lllll@{}}
\toprule
\textbf{Date}      & \multicolumn{2}{c}{$\boldsymbol{H_0: \alpha \leq 1}$}                                                 & \multicolumn{2}{c}{$\boldsymbol{H_0: \alpha \leq 2}$}                                                          \\ \midrule
\textbf{}          & \multicolumn{1}{c}{$\chi^2$\textit{\textbf{ Statistic}}} & \multicolumn{1}{c}{\textit{\textbf{p-value}}} & \multicolumn{1}{c}{$\chi^2$\textit{\textbf{ Statistic}}} & \multicolumn{1}{c}{\textit{\textbf{p-value}}} \\ \cmidrule(l){2-5} 
\textit{7/1/2021}  & 79.62                                              & $4.54 \times 10^{-19}  $                           & 30.90                                              & $2.72 \times 10^{-8}  $                               \\
\textit{10/1/2021} & $4.70 \times 10^{-5}$                                          & 0.99                                 & 0                                                  & 1                                             \\
\textit{1/1/2022}  & $3.36 \times 10^{-7}   $                                       & 1                                    & 0                                                  & 1                                             \\
\textit{4/1/2022}  & 0                                                  & 1                                    & 0                                                  & 1                                             \\
\textit{7/1/2022}  & 41.48                                              & $1.19 \times 10^{-10} $                           & 1.92                                               & 0.17                                          \\
\textit{10/1/2022} & 74.65                                              & $5.63\times 10^{-18}   $                         & 11.35                                              & $7.53 \times 10^{-4}     $                                 \\
\textit{1/1/2023}  & 111.87                                             & $3.82 \times 10^{-26}   $                          & 28.45                                              & $9.62 \times 10^{-8} $                                    \\ \bottomrule
\end{tabular}
\end{table}

\subsection*{Case 2: Monkeypox Outbreak 2022: France}
\subsubsection*{Data Description and Analysis Plan}
We also used the monkeypox (mpox) data set on Our World In Data \cite{owidmonkeypox} to assess the performance of new discrete composite models. The numbers of reported cases by days in France from May 19st, 2022 to January 1st, 2023 were included in our analysis. The time series for daily cases count is presented in Figure \ref{fig:cs_sin}. Unlike the COVID-19 data set we chose in case 1, this data set contains a considerable amount of zeros. We first analyze the reported cases from May 19st, 2022 to July 1st, 2022. Then, we update our analysis by incorporating the cases from next month. Sequentially, we were able to obtain the analysis results on seven particular dates: 7/1/2022, 8/1/2021, 9/1/2022, 10/1/2022, 11/1/2022, 12/1/2022, and 1/1/2023. 

    \begin{figure}[pos=H]

\includegraphics[width=12cm]{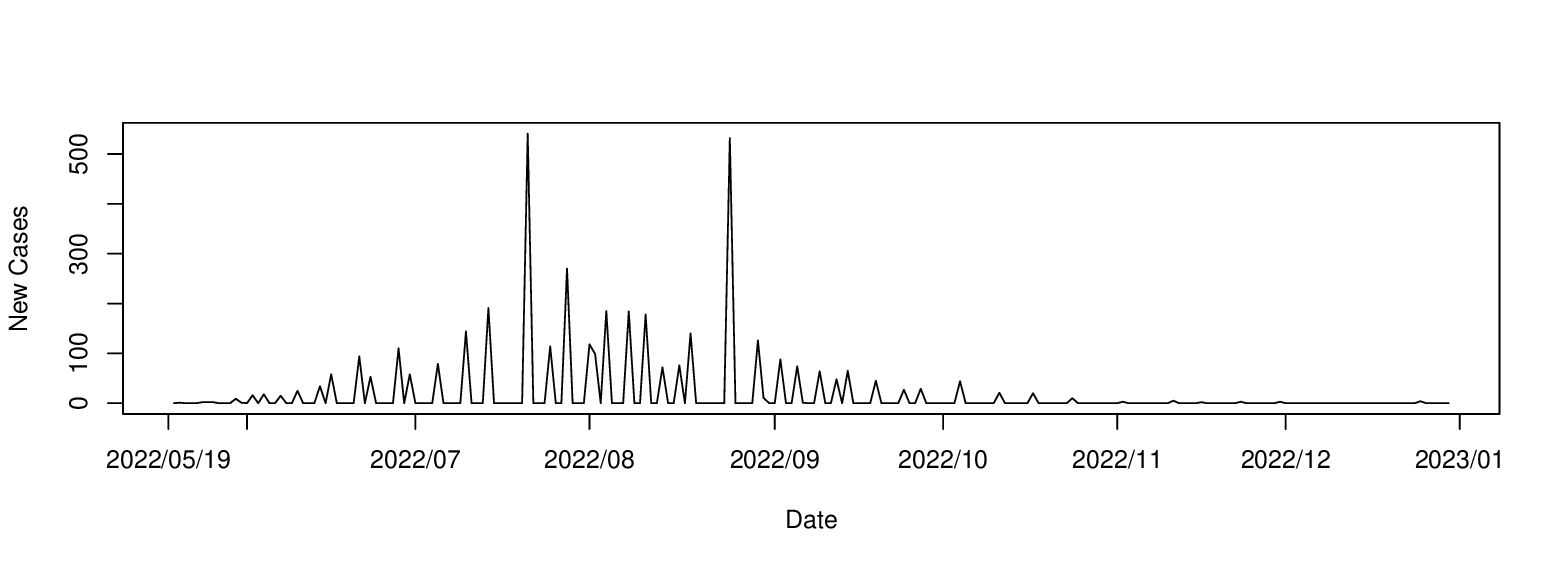}
\centering
\caption{The mpox cases over time in France from 5/19/2022 to 01/01/2023}
\label{fig:6}
\end{figure}

\subsubsection*{Results}
\begin{figure}[pos=H]
    \includegraphics[width=12cm]{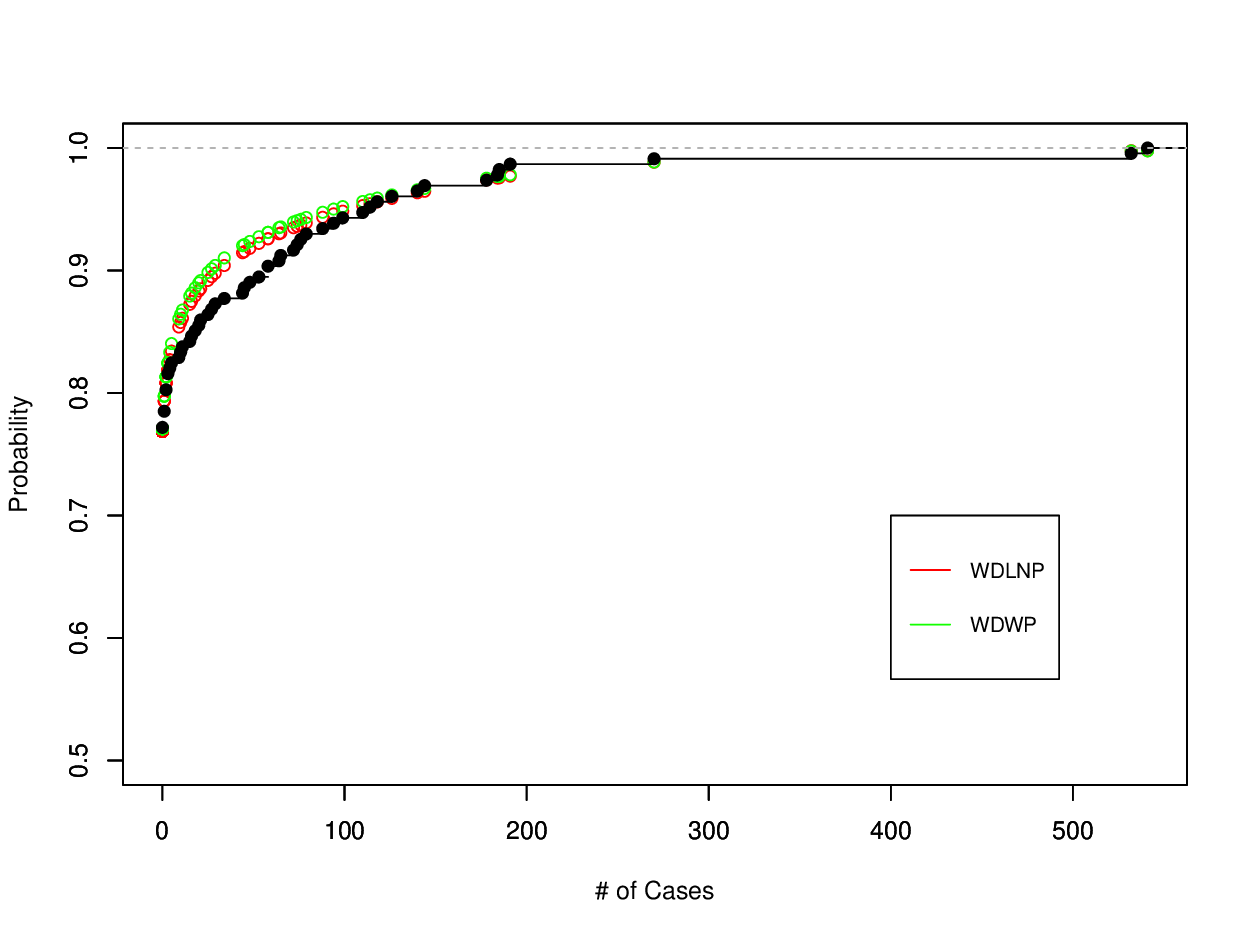}
\centering
\caption{The empirical CDF of mpox cases in France from 5/19/2022 to 1/1/2023 with the fitted CDF of two composite distributions}
\label{fig:7}
\end{figure}

Table \ref{table:4} shows the AIC values for six different models on seven particular dates. Models with the lowest AIC values were highlighted in red. For the dates 7/1/2022, 9/1/2022, 10/1/2022, and 11/1/2022, the discrete composite models demonstrated better performance compared to well-known models. Figure \ref{fig:7} presents the estimated CDF plot with two discrete composite models with the empirical CDF when we included all the daily reported count cases from 5/19/2022 to 1/1/2023. Both models demonstrated reasonably good performance when to fitting to upper tail. However, both lognormal and Weibull distribution cannot capture the lower tail features of the data well based on Figure \ref{fig:7}. This suggest better choices of the head distributions for the discrete composite models might exist when fitting to the data set. 

\begin{table}[pos = H]
\caption{AIC Values for Different Models (Mpox Cases in France: 5/19/2022-1/1/2023)}
\label{table:4}
\begin{tabular}{llllllll}
\hline
                 & \textbf{Poisson} & \textbf{ZIP}\textsuperscript{2} & \textbf{NB}\textsuperscript{3}                     & \textbf{ZINB}\textsuperscript{4} & \textbf{WDLNP}\textsuperscript{5}                  & \textbf{WDWP}\textsuperscript{6}                   \\ \hline
\textbf{Date}\textsuperscript{1}    &                  &               &                                 &                &              &                                                               \\ \hline
\textit{7/1/2022} & 1629.815         & 681.9589     & 200.2278 & 201.1567          & \color{red}200.1003                        & 209.1702      \\ \hline
\textit{8/1/2022} & 7512.623         & 2985.267     & \color{red}340.0716 & 342.4696               & 341.7540                         & 342.6394      \\ \hline
\textit{9/1/2022} & 12538.09         & 4367.755     & 514.1144                         & 516.1153           & \color{red} 512.073  & 514.2812      \\ \hline
\textit{10/1/2022} & 14133.59         & 4906.265     & 647.5542                         & 649.5543           & \color{red} 643.7592 & 646.7542      \\ \hline
\textit{11/1/2022} & 15626.52         & 5267.969     & 715.6079                         & 717.608           & \color{red} 711.1758 & 713.905       \\ \hline
\textit{12/1/2022} & 16918.46         & 5880.85      & 766.7535                         & \color{red} 760.442              & 763.0148 & 765.2526      \\ \hline
\textit{1/1/2023} & 18088.32         & 6173.418     & 801.2655                         & \color{red} 796.9182                  & 797.7298 & 799.452       \\ \hline
\end{tabular}\\
\noindent{ \footnotesize{\textsuperscript{1} The data was included from 5/19/2022 until this date.}}\\
	\noindent{ \footnotesize{\textsuperscript{2} Zero-inflated Poisson}}\\
 \noindent{ \footnotesize{\textsuperscript{3} Negative binomial}}\\
  \noindent{ \footnotesize{\textsuperscript{4} Zero-inflated negative binomial}}\\

  \noindent{ \footnotesize{\textsuperscript{5} Weighted discrete lognormal-Pareto}}\\
  \noindent{ \footnotesize{\textsuperscript{6} Weighted discrete Weibull-Pareto}}
\end{table}

    \begin{figure}[pos=H]

\includegraphics[width=12cm]{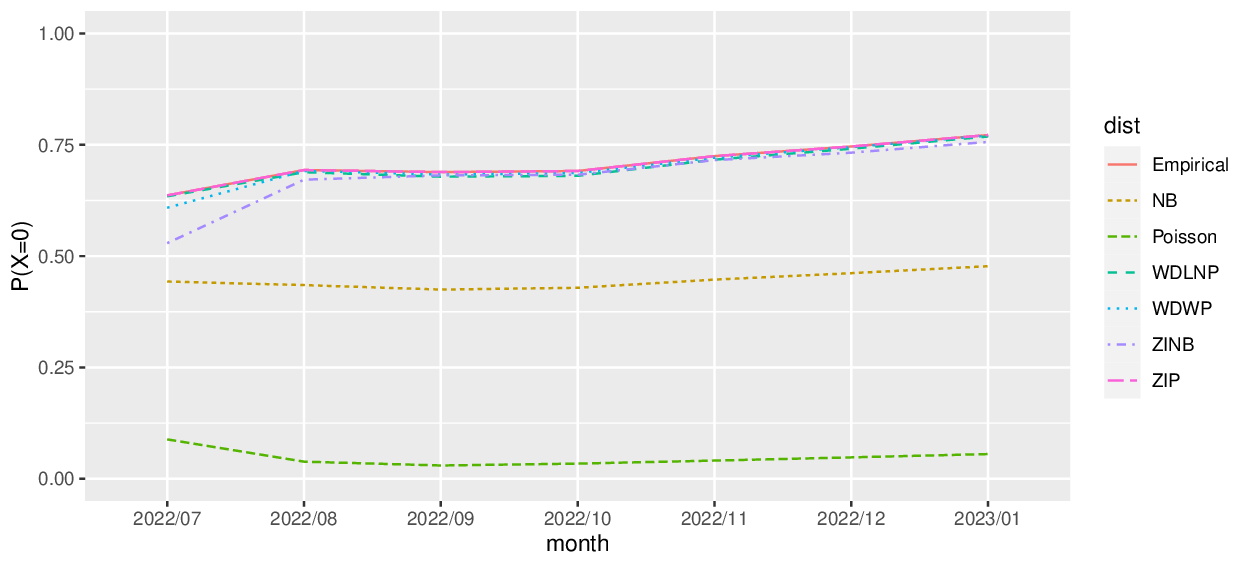}
\centering
\caption{The estimated probability of $0$ cases for different models}
\label{fig:8}
\end{figure}

Figure \ref{fig:8} shows the estimated probability for zero cases for different models when fitting to the mpox data. In terms of the ability to capture the probability of zero cases, our model can demonstrate comparable performance to that of zero-inflated models without adding a parameter for modeling the probability of zero cases.

        \begin{figure}[pos=H]
        
\includegraphics[width=12cm]{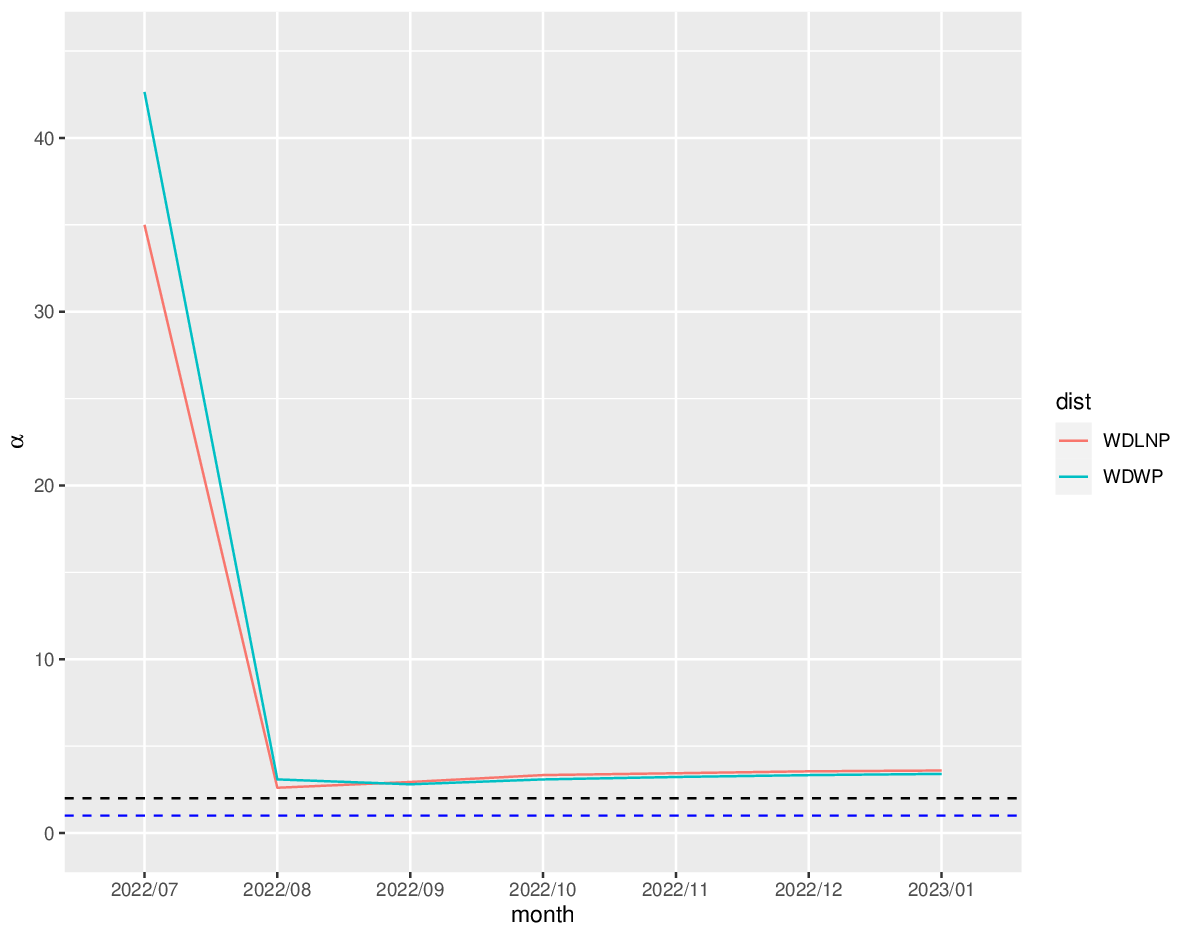}
\centering
\caption{The change of $\alpha$ estimates over time for two composite models when fitting to the mpox data from France}
\label{fig:9}
\end{figure}

Figure \ref{fig:9} demonstrates the change of $\alpha$ estimates over time for two discrete composite models. Again, the black horizontal dashed line stands for $\alpha = 2$, and the blue horizontal dashed line represents $\alpha = 1$. Note for both models, the $\alpha$ estimates are greater than $2$ during 7/1/2021 and 01/01/2022. Table \ref{table:5} and \ref{table:6} shows the results of the hypothesis testing for the tail index parameter $\alpha$ for two models. We notice that only for the date 8/1/2022, the null hypothesis of $\alpha=1$ should not be rejected for both models at the significance level of 0.05. This suggests that the estimated expected counts of cases for both models are finite for most of the scenarios. However, during the whole observation period, $\alpha = 2$ should not be rejected for both models at the significance level of 0.05. This the counts were possibly associated with infinite variance, even the estimates of $\alpha$ are over 2 for the whole time. 

Similar to the analysis we did for the COVID-19 cases in Singapore, we can view the results from Figure \ref{fig:9}, Table \ref{table:5}, and Table \ref{table:6} with the time series plot in Figure \ref{fig:6} for the mpox cases in France. Notice we reject both $\alpha \leq 1$ and $\alpha \leq 2$ on 7/1/2022 as the estimated mean and variance are expected to be finite. This implies mpox should not be considered as a major public health concern at that time. On 8/1/2022, the $\alpha$ estimates for both models drops to nearly 2 and the LRT suggests that the null hypothesis of infinite mean and variance should not be rejected. This makes sense since we saw the large number of reported cases in July and local health agencies should plan for extremely high counts in mpox cases. However, even though some new cases were reported in August or September, they did not cause drastic changes in the estimate of $\alpha$ for both models. The $\alpha$ estimates for the dates after 8/1/2022 stayed beyond 2 until 1/1/2023. This suggests the mpox outbreak in France was in a controllable situation after the first wave of reported cases. 

    \begin{table}[pos=H]
        \caption{The LRT test results for tail index $\alpha$ with WDLNP model when fitting to mpox data in France}
        \label{table:5}
\begin{tabular}{lllll}
\hline
\textbf{Date}      & \multicolumn{2}{c}{$\boldsymbol{H_0: \alpha \leq 1}$}                                                & \multicolumn{2}{c}{$\boldsymbol{H_0: \alpha \leq 2}$}                                                 \\ \hline
\textbf{}          & \multicolumn{1}{c}{\textit{\textbf{$\chi^2$ Statistic}}} & \multicolumn{1}{c}{\textit{\textbf{p-value}}} & \multicolumn{1}{c}{\textit{\textbf{$\chi^2$ Statistic}}} & \multicolumn{1}{c}{\textit{\textbf{p-value}}} \\ \cline{2-5} 
\textit{7/1/2022}  & 5.77                                           & 0.02                                    & 3.60                                           & 0.06                                    \\
\textit{8/1/2022}  & 2.57                                           & 0.11                                     & 0.15                                          & 0.70                                    \\
\textit{9/1/2022}  & 4.93                                           & 0.03                                    & 0.94                                          & 0.33                                      \\
\textit{10/1/2022} & 5.94                                           & 0.02                                    & 1.10                                           & 0.29                                     \\
\textit{11/1/2022} & 6.15                                           & 0.01                                    & 1.10                                           & 0.29                                     \\
\textit{12/1/2022} & 6.26                                           & 0.01                                    & 1.09                                           & 0.30                                     \\
\textit{1/1/2023}  & 6.29                                            & 0.01                                    & 1.03                                           & 0.31                  \\
\hline
\end{tabular}
\end{table}

\begin{table}[pos = H]
\caption{The LRT test results for tail index $\alpha$ with WDWP model when fitting to mpox data in France}
\label{table:6}
\begin{tabular}{ccccc}
\hline
\textbf{Date}      & \multicolumn{2}{c}{$\boldsymbol{H_0: \alpha \leq 1}$}        & \multicolumn{2}{c}{$\boldsymbol{H_0: \alpha \leq 2}$}         \\ \hline
\textit{\textbf{}} & \textit{\textbf{$\chi^2$ Statistic}} & \textit{\textbf{p-value}} & \textit{\textbf{$\chi^2$ Statistic}} & \textit{\textbf{p-value}} \\ \cline{2-5} 
\textit{7/1/2022}  & 4.74                       & 0.03                 & 2.54                      & 0.11                 \\
\textit{8/1/2022}  & 2.43                       & 0.12                 & 0.29                      & 0.59                 \\
\textit{9/1/2022}  & 4.36                       & 0.04                & 0.72                      & 0.40                 \\
\textit{10/1/2022} & 3.85                       & 0.05                & 1.10                       & 0.29                 \\
\textit{11/1/2022} & 5.84                       & 0.02                & 1.19                       & 0.28                 \\
\textit{12/1/2022} & 5.99                       & 0.01               & 1.19                       & 0.27                 \\
\textit{1/1/2023}  & 6.04                      & 0.01                & 1.17                       & 0.28   \\
\hline
\end{tabular}
\end{table}

\section{Concluding Remarks}
\label{Section:6}
In this paper, we introduced the new discrete composite distributions with Pareto tails. We created two novel discrete composite distributions named weighted discrete lognormal-Pareto distribution (WDLNP) and weighted discrete Weibull-Pareto distribution (WDWP). Some properties about the new distributions were derived. In addition, we developed a testing procedure to test the hypotheses of infinite mean and infinite variance of the discrete composite models. Two novel discrete composite models can provide very good performance when fitting to real infectious disease data. We also justified the capability of these two models when fitting to the data with excess zeros by using real data. 

The findings from the real data application are interesting. As Poisson distribution is widely used in modeling the cases of infectious diseases, it demonstrated poor performance when fitting to both real data sets we selected. Like some previous literature suggested \cite{cohen_covid-19_2022}, Poisson distribution does not perform well when the data is characterized with heavy-tailed features. As a comparison, our models demonstrated great and consistent performance when fitting to two different infectious diseases data sets due to the Pareto tail we selected. 
In addition, the estimated Pareto tail index parameter of these models can provide us a way to monitor the severity of the an infectious disease in terms of reported cases. As we showed in two real data examples, we were able to sequentially update our models and testing results by incorporating new cases reported. By tracing the change of tail index parameter and hypothesis testing results, we were able to provide a way to monitor the progression of a pandemic such as COVID-19. 

\appendix

\printcredits

\bibliographystyle{model1-num-names}
%\bibliographystyle{cas-model2-names}

\bibliography{references}

\begin{thebibliography}{27}
\expandafter\ifx\csname natexlab\endcsname\relax\def\natexlab#1{#1}\fi
\providecommand{\url}[1]{\texttt{#1}}
\providecommand{\href}[2]{#2}
\providecommand{\path}[1]{#1}
\providecommand{\DOIprefix}{doi:}
\providecommand{\ArXivprefix}{arXiv:}
\providecommand{\URLprefix}{URL: }
\providecommand{\Pubmedprefix}{pmid:}
\providecommand{\doi}[1]{\href{http://dx.doi.org/#1}{\path{#1}}}
\providecommand{\Pubmed}[1]{\href{pmid:#1}{\path{#1}}}
\providecommand{\bibinfo}[2]{#2}
\ifx\xfnm\relax \def\xfnm[#1]{\unskip,\space#1}\fi
%Type = Article
\bibitem[{Brown et~al.(2021)Brown, Cohen, Tang, and Yam}]{brown_taylors_2021}
\bibinfo{author}{M.~Brown}, \bibinfo{author}{J.~E. Cohen}, \bibinfo{author}{C.-F. Tang}, \bibinfo{author}{S.~C.~P. Yam},
\newblock \bibinfo{title}{Taylor’s law of fluctuation scaling for semivariances and higher moments of heavy-tailed data},
\newblock \bibinfo{journal}{Proceedings of the National Academy of Sciences} \bibinfo{volume}{118} (\bibinfo{year}{2021}) \bibinfo{pages}{e2108031118}. \bibinfo{note}{Publisher: Proceedings of the National Academy of Sciences}.
%Type = Article
\bibitem[{Cohen et~al.(2022)Cohen, Davis, and Samorodnitsky}]{cohen_covid-19_2022}
\bibinfo{author}{J.~E. Cohen}, \bibinfo{author}{R.~A. Davis}, \bibinfo{author}{G.~Samorodnitsky},
\newblock \bibinfo{title}{{COVID}-19 cases and deaths in the {United} {States} follow {Taylor}’s law for heavy-tailed distributions with infinite variance},
\newblock \bibinfo{journal}{Proceedings of the National Academy of Sciences} \bibinfo{volume}{119} (\bibinfo{year}{2022}) \bibinfo{pages}{e2209234119}. \bibinfo{note}{Publisher: Proceedings of the National Academy of Sciences}.
%Type = Article
\bibitem[{Beare and Toda(2020)}]{beare_emergence_2020}
\bibinfo{author}{B.~K. Beare}, \bibinfo{author}{A.~A. Toda},
\newblock \bibinfo{title}{On the emergence of a power law in the distribution of {COVID}-19 cases},
\newblock \bibinfo{journal}{Physica D. Nonlinear Phenomena} \bibinfo{volume}{412} (\bibinfo{year}{2020}) \bibinfo{pages}{132649}.
%Type = Article
\bibitem[{Blasius(2020)}]{blasius_power-law_2020}
\bibinfo{author}{B.~Blasius},
\newblock \bibinfo{title}{Power-law distribution in the number of confirmed {COVID}-19 cases},
\newblock \bibinfo{journal}{Chaos} \bibinfo{volume}{30} (\bibinfo{year}{2020}) \bibinfo{pages}{093123}.
%Type = Article
\bibitem[{Cooray and Ananda(2005)}]{ananda2005}
\bibinfo{author}{K.~Cooray}, \bibinfo{author}{M.~M.~A. Ananda},
\newblock \bibinfo{title}{Modeling actuarial data with a composite lognormal-pareto model},
\newblock \bibinfo{journal}{Scandinavian Actuarial Journal} \bibinfo{volume}{2005} (\bibinfo{year}{2005}) \bibinfo{pages}{321--334}.
%Type = Article
\bibitem[{Scollnik(2007)}]{scollnik_composite_2007}
\bibinfo{author}{D.~P.~M. Scollnik},
\newblock \bibinfo{title}{On composite lognormal-{Pareto} models},
\newblock \bibinfo{journal}{Scandinavian Actuarial Journal} \bibinfo{volume}{2007} (\bibinfo{year}{2007}) \bibinfo{pages}{20--33}.
%Type = Article
\bibitem[{Scollnik and Sun(2012)}]{Scollnik2012ModelingWW}
\bibinfo{author}{D.~P.~M. Scollnik}, \bibinfo{author}{C.~Sun},
\newblock \bibinfo{title}{Modeling with weibull-pareto models},
\newblock \bibinfo{journal}{North American Actuarial Journal} \bibinfo{volume}{16} (\bibinfo{year}{2012}) \bibinfo{pages}{260 -- 272}.
%Type = Article
\bibitem[{Brazauskas and Kleefeld(2016)}]{brazauskas_modeling_2016}
\bibinfo{author}{V.~Brazauskas}, \bibinfo{author}{A.~Kleefeld},
\newblock \bibinfo{title}{Modeling {Severity} and {Measuring} {Tail} {Risk} of {Norwegian} {Fire} {Claims}},
\newblock \bibinfo{journal}{North American Actuarial Journal} \bibinfo{volume}{20} (\bibinfo{year}{2016}) \bibinfo{pages}{1--16}.
%Type = Article
\bibitem[{Grün and Miljkovic(2019)}]{grun2019}
\bibinfo{author}{B.~Grün}, \bibinfo{author}{T.~Miljkovic},
\newblock \bibinfo{title}{Extending composite loss models using a general framework of advanced computational tools},
\newblock \bibinfo{journal}{Scandinavian Actuarial Journal} \bibinfo{volume}{2019} (\bibinfo{year}{2019}) \bibinfo{pages}{642--660}.
%Type = Article
\bibitem[{Liu and Ananda(2022{\natexlab{a}})}]{liu_analyzing_2022}
\bibinfo{author}{B.~Liu}, \bibinfo{author}{M.~M.~A. Ananda},
\newblock \bibinfo{title}{Analyzing insurance data with an exponentiated composite {Inverse}-{Gamma} {Pareto} {Model}},
\newblock \bibinfo{journal}{Communications in Statistics - Theory and Methods}  (\bibinfo{year}{2022}{\natexlab{a}}).
%Type = Article
\bibitem[{Liu and Ananda(2022{\natexlab{b}})}]{liu_generalized_2022}
\bibinfo{author}{B.~Liu}, \bibinfo{author}{M.~M.~A. Ananda},
\newblock \bibinfo{title}{A {Generalized} {Family} of {Exponentiated} {Composite} {Distributions}},
\newblock \bibinfo{journal}{Mathematics} \bibinfo{volume}{10} (\bibinfo{year}{2022}{\natexlab{b}}) \bibinfo{pages}{1895}. \bibinfo{note}{Number: 11 Publisher: Multidisciplinary Digital Publishing Institute}.
%Type = Article
\bibitem[{Mutali and Vernic(2020)}]{mutali_composite_2020}
\bibinfo{author}{S.~Mutali}, \bibinfo{author}{R.~Vernic},
\newblock \bibinfo{title}{On the composite {Lognormal}–{Pareto} distribution with uncertain threshold},
\newblock \bibinfo{journal}{Communications in Statistics - Simulation and Computation} \bibinfo{volume}{0} (\bibinfo{year}{2020}) \bibinfo{pages}{1--17}. \bibinfo{note}{Publisher: Taylor \& Francis}.
%Type = Article
\bibitem[{Deng and Aminzadeh(2019)}]{deng_bayesian_2019}
\bibinfo{author}{M.~Deng}, \bibinfo{author}{M.~S. Aminzadeh},
\newblock \bibinfo{title}{Bayesian predictive analysis for {Weibull}-{Pareto} composite model with an application to insurance data},
\newblock \bibinfo{journal}{Communications in Statistics - Simulation and Computation} \bibinfo{volume}{0} (\bibinfo{year}{2019}) \bibinfo{pages}{1--27}.
%Type = Article
\bibitem[{Aminzadeh and Deng(2019)}]{ig_pareto}
\bibinfo{author}{M.~S. Aminzadeh}, \bibinfo{author}{M.~Deng},
\newblock \bibinfo{title}{Bayesian predictive modeling for inverse gamma-pareto composite distribution},
\newblock \bibinfo{journal}{Communications in Statistics - Theory and Methods} \bibinfo{volume}{48} (\bibinfo{year}{2019}) \bibinfo{pages}{1938--1954}.
%Type = Article
\bibitem[{Nadarajah(2005)}]{exp_pareto}
\bibinfo{author}{S.~Nadarajah},
\newblock \bibinfo{title}{Exponentiated pareto distributions},
\newblock \bibinfo{journal}{Statistics} \bibinfo{volume}{39} (\bibinfo{year}{2005}) \bibinfo{pages}{255--260}.
%Type = Article
\bibitem[{Calderín-Ojeda(2018)}]{calderin-ojeda_note_2018}
\bibinfo{author}{E.~Calderín-Ojeda},
\newblock \bibinfo{title}{A {Note} on {Parameter} {Estimation} in the {Composite} {Weibull}–{Pareto} {Distribution}},
\newblock \bibinfo{journal}{Risks} \bibinfo{volume}{6} (\bibinfo{year}{2018}) \bibinfo{pages}{11}.
%Type = Article
\bibitem[{Cooray(2009)}]{cooray_weibullpareto_2009}
\bibinfo{author}{K.~Cooray},
\newblock \bibinfo{title}{The {Weibull}–{Pareto} {Composite} {Family} with {Applications} to the {Analysis} of {Unimodal} {Failure} {Rate} {Data}},
\newblock \bibinfo{journal}{Communications in Statistics - Theory and Methods} \bibinfo{volume}{38} (\bibinfo{year}{2009}) \bibinfo{pages}{1901--1915}.
%Type = Article
\bibitem[{Cooray et~al.(2010)Cooray, Gunasekera, and Ananda}]{cooray_weibull_2010}
\bibinfo{author}{K.~Cooray}, \bibinfo{author}{S.~Gunasekera}, \bibinfo{author}{M.~Ananda},
\newblock \bibinfo{title}{Weibull and inverse {Weibull} composite distribution for modeling reliability data},
\newblock \bibinfo{journal}{MAS} \bibinfo{volume}{5} (\bibinfo{year}{2010}) \bibinfo{pages}{109--115}.
%Type = Article
\bibitem[{Liu and Ananda(2023)}]{liu_new_2023}
\bibinfo{author}{B.~Liu}, \bibinfo{author}{M.~M.~A. Ananda},
\newblock \bibinfo{title}{A {New} {Insight} into {Reliability} {Data} {Modeling} with an {Exponentiated} {Composite} {Exponential}-{Pareto} {Model}},
\newblock \bibinfo{journal}{Applied Sciences} \bibinfo{volume}{13} (\bibinfo{year}{2023}) \bibinfo{pages}{645}. \bibinfo{note}{Number: 1 Publisher: Multidisciplinary Digital Publishing Institute}.
%Type = Article
\bibitem[{Chakraborty(2015)}]{chakraborty_generating_2015}
\bibinfo{author}{S.~Chakraborty},
\newblock \bibinfo{title}{Generating discrete analogues of continuous probability distributions-{A} survey of methods and constructions},
\newblock \bibinfo{journal}{Journal of Statistical Distributions and Applications} \bibinfo{volume}{2} (\bibinfo{year}{2015}) \bibinfo{pages}{6}.
%Type = Article
\bibitem[{Kemp(2004)}]{kemp_classes_2004}
\bibinfo{author}{A.~W. Kemp},
\newblock \bibinfo{title}{Classes of {Discrete} {Lifetime} {Distributions}},
\newblock \bibinfo{journal}{Communications in Statistics - Theory and Methods} \bibinfo{volume}{33} (\bibinfo{year}{2004}) \bibinfo{pages}{3069--3093}. \bibinfo{note}{Publisher: Taylor \& Francis}.
%Type = Article
\bibitem[{Roy(2003)}]{roy_discrete_2003}
\bibinfo{author}{D.~Roy},
\newblock \bibinfo{title}{The {Discrete} {Normal} {Distribution}},
\newblock \bibinfo{journal}{Communications in Statistics - Theory and Methods} \bibinfo{volume}{32} (\bibinfo{year}{2003}) \bibinfo{pages}{1871--1883}.
%Type = Article
\bibitem[{Abu~Bakar et~al.(2015)Abu~Bakar, Hamzah, Maghsoudi, and Nadarajah}]{Bak15}
\bibinfo{author}{S.~Abu~Bakar}, \bibinfo{author}{N.~Hamzah}, \bibinfo{author}{M.~Maghsoudi}, \bibinfo{author}{S.~Nadarajah},
\newblock \bibinfo{title}{{Modeling loss data using composite models}},
\newblock \bibinfo{journal}{Insurance: Mathematics and Economics} \bibinfo{volume}{61} (\bibinfo{year}{2015}) \bibinfo{pages}{146--154}.
%Type = Article
\bibitem[{Almetwally et~al.(2022)Almetwally, Abdo, Hafez, Jawa, Sayed-Ahmed, and Almongy}]{almetwally_new_2022}
\bibinfo{author}{E.~M. Almetwally}, \bibinfo{author}{D.~A. Abdo}, \bibinfo{author}{E.~Hafez}, \bibinfo{author}{T.~M. Jawa}, \bibinfo{author}{N.~Sayed-Ahmed}, \bibinfo{author}{H.~M. Almongy},
\newblock \bibinfo{title}{The new discrete distribution with application to {COVID}-19 {Data}},
\newblock \bibinfo{journal}{Results in Physics} \bibinfo{volume}{32} (\bibinfo{year}{2022}) \bibinfo{pages}{104987}.
%Type = Article
\bibitem[{Almetwally et~al.(2023)Almetwally, Dey, and Nadarajah}]{almetwally_overview_2023}
\bibinfo{author}{E.~M. Almetwally}, \bibinfo{author}{S.~Dey}, \bibinfo{author}{S.~Nadarajah},
\newblock \bibinfo{title}{An {Overview} of {Discrete} {Distributions} in {Modelling} {COVID}-19 {Data} {Sets}},
\newblock \bibinfo{journal}{Sankhya A} \bibinfo{volume}{85} (\bibinfo{year}{2023}) \bibinfo{pages}{1403--1430}.
%Type = Book
\bibitem[{Burnham and Anderson(2002)}]{burnham}
\bibinfo{author}{K.~P. Burnham}, \bibinfo{author}{D.~R. Anderson}, \bibinfo{title}{Model Selection and Multimodel Inference}, \bibinfo{edition}{2} ed., \bibinfo{publisher}{Springer-Verlag}, \bibinfo{year}{2002}.
%Type = Article
\bibitem[{Mathieu et~al.(2022)Mathieu, Spooner, Dattani, Ritchie, and Roser}]{owidmonkeypox}
\bibinfo{author}{E.~Mathieu}, \bibinfo{author}{F.~Spooner}, \bibinfo{author}{S.~Dattani}, \bibinfo{author}{H.~Ritchie}, \bibinfo{author}{M.~Roser},
\newblock \bibinfo{title}{Mpox (monkeypox)},
\newblock \bibinfo{journal}{Our World in Data}  (\bibinfo{year}{2022}). \bibinfo{note}{Https://ourworldindata.org/monkeypox}.

\end{thebibliography}


\end{document}